\documentclass[prl, aps, twocolumn, groupedaddress, superscriptaddress,
               showpacs, floatfix, preprintnumbers]{revtex4-1}
\usepackage[multidot]{grffile}  
\usepackage{amsmath,amssymb,multirow,epsfig,bm}

\newcommand{\vect}[1]{\bm{#1}}
\newcommand{\uvect}[1]{\bm{\hat{#1}}}
\newcommand{\abs}[1]{\lvert#1\rvert}
\newcommand{\I}{\mathrm{i}}
\renewcommand{\d}{\mathrm{d}}

\newcommand{\beq}{\begin{equation}}
\newcommand{\eeq}{\end{equation}}
\newcommand{\bea}{\begin{eqnarray}}
\newcommand{\eea}{\end{eqnarray}}

\newcommand{\kF}{k_{\textrm{F}}}
\newcommand{\fm}{\;\textrm{fm}}
\newcommand{\MeV}{\;\textrm{MeV}}

\newcommand{\exclude}[1]{}

\begin{document}
\preprint{NT@UW-16-06}

\title{Vortex pinning and dynamics in the neutron star crust}

\author{Gabriel Wlaz\l{}owski}
\email{gabrielw@if.pw.edu.pl}
\affiliation{Faculty of Physics, %
  Warsaw University of Technology, %
  Ulica Koszykowa 75, 00--662 Warsaw, POLAND}
\affiliation{Department of Physics, %
  University of Washington, %
  Seattle, Washington 98195--1560, USA}

\author{Kazuyuki Sekizawa}
\email{sekizawa@if.pw.edu.pl}
\affiliation{Faculty of Physics, Warsaw University of Technology, %
  Ulica Koszykowa 75, 00--662 Warsaw, POLAND}

\author{Piotr Magierski},
\email{magiersk@if.pw.edu.pl}
\affiliation{Faculty of Physics, Warsaw University of Technology, %
  Ulica Koszykowa 75, 00--662 Warsaw, POLAND}
\affiliation{Department of Physics, University of Washington, Seattle, %
  Washington 98195--1560, USA}

\author{Aurel Bulgac}, 
\email{bulgac@uw.edu}
\affiliation{Department of Physics, University of Washington, Seattle, %
  Washington 98195--1560, USA}

\author{Michael McNeil Forbes}
\email{michael.forbes@wsu.edu}
\affiliation{Department of Physics \& Astronomy, Washington State University, %
  Pullman, WA 98164--2814, USA}
\affiliation{Department of Physics, University of Washington, Seattle, %
  Washington 98195--1560, USA}

\begin{abstract} 
  The nature of the interaction between superfluid vortices and the neutron
  star crust, conjectured by Anderson and Itoh in 1975 to be at the heart
  vortex creep and the cause of glitches, has been a long-standing question in
  astrophysics.  
  Using a
  qualitatively new approach, we follow the dynamics as superfluid vortices 
  move in response to the presence of ``nuclei'' (nuclear defects in the crust).  
  The resulting motion is perpendicular to the force, similar to the motion of 
  a spinning top when pushed. We show that nuclei repel vortices in the neutron
  star crust, and characterize the 
  force as a function of the vortex-nucleus separation. 
\end{abstract}

\date{\today}

\pacs{26.60.Gj, 03.75.Kk, 67.10.Jn, 67.25.dk }

\maketitle
\paragraph{Introduction}
Pulsar glitches, sudden increases in the pulsation frequency first observed in
1969~\cite{Radhakrishnan:1969, Reichley:1969}, provide one of the few
observable probes into the interior of neutron stars~\cite{LattimerPrakash}.
Although many models have been proposed, the origin of large glitches remains a
mystery.  The current picture, proposed in 1975 by Anderson and
Itoh~\cite{AndersonItoh}, is that the quantized vortices in the superfluid
interior of a neutron star store a significant amount of angular momentum.  As
these vortices ``creep'' through the crust, they transfer this angular momentum
to the crust. Glitches result from a catastrophic release of pinned
vorticity~\cite{ModelsGlitch} that suddenly changes the pulsation rate.

This scenario involves two critical ingredients: the trigger mechanism for the
catastrophic release (not considered here) and the vortex-``nucleus''
interaction. (By ``nucleus'' we mean nucleilike objects embedded in a neutron
superfluid as is expected in the crust of neutron stars.)
The interaction can, in principle, be derived from a microscopic theory. However,
despite considerable theoretical effort, even its sign remains uncertain.
Until now, the force was evaluated by comparing (free) energies extracted from
different static calculations: a vortex passing through a nucleus, a vortex
and nucleus separated by an infinite distance, or an interstitial vortex
between two neighboring nuclei~\cite{Pizzochero2, Donati, Donati2, Donati3,
  Seveso, Avogadro1, Avogadro2}.  As pointed out in
Ref.~\cite{Pizzochero1}, this approach only computes the pinning energy, and is
not able to extract the full information about the vortex-nucleus interaction.

Despite these efforts, there is still no agreement about whether pinned or
unpinned configurations are preferred. The problem is that subtracting two
large energies arising from many contributions, typically of order $10^4\MeV$,
results in a tiny difference of order $1\MeV$.
Symmetry-unrestricted calculations are challenging, and only by imposing axial
symmetry has the required $1\MeV$
accuracy been achieved~\cite{Avogadro2, COCG}.  Moreover, the difference is
extremely sensitive to quantum shell effects that are not present in
semiclassical simulations~\cite{Pizzochero2, Donati, Donati2, Donati3}, and is
sensitive to the particle number or background density which varies from one
configuration to the next. (See~\cite{Avogadro2} for extensive discussion.)

Here we show that nuclei repel superfluid vortices and characterize the
vortex-nucleus interaction within dynamical simulations as suggested
in~\cite{BulgacPinning}. The sign of the force can be unambiguously determined
by looking at the vortex motion (see the movie demonstrating the
response of classical gyroscope when pushed  in the Supplemental Material~\cite{Supplemental}).  
In 3D
dynamical simulations, all relevant degrees of freedom of the vortex-nucleus
system are active, and the behavior provides valuable insight for building
effective theories of the vortex-nucleus system. Following~\cite{BLink}, an
effective hydrodynamic description can be formulated (see also~\cite{Quantvor,
  Epstein, Baym, Antonelli})
\begin{equation}
  T\dfrac{\partial^2\vect{r}}{\partial z^2} 
  + \rho_s\vect{\kappa}\times\left(
    \dfrac{\partial\vect{r}}{\partial t} - \vect{v}_s\right) 
  + \vect{f}_{\text{VN}}
  =0,\label{eqn:vmotion}
\end{equation}
where $\vect{r}$
is the position of the vortex core. The first term is tension force as the vortex
is bent, characterized by coefficient $T$.
The second term corresponds to Magnus force where
$\vect{\kappa} = 2\pi \hbar \uvect{l}/2m_n$
is the circulation which points along the vortex, $\rho_s=m_n n$
is mass density while $n$
is number density of superfluid neutron background, $m_n$
is neutron mass, and $\vect{v}_s$
is the velocity of any ambient flow in the background superfluid density.  The
last contribution $\vect{f}_{\text{VN}}$
defines the vortex-nucleus interaction in terms of the force per unit
length. Clearly the pinning energy alone does not provide sufficient
information to describe the motion.  In addition, all existing calculations
assume that the vortices form straight lines, and thus do not reveal
information about the tension.  It is demonstrated in~\cite{BLink} that pinning
occurs irrespective of the sign of the vortex-nucleus interaction when
$v_s<v_c\sim s^{1/2}F_m/\rho_s\kappa a$,
where $F_m$
is set by maximum magnitude of $\vect{f}_{\text{VN}}$,
$s=F_m/T$,
and $a$
is set by range of vortex-nucleus interaction.  In this work we extract all
effective quantities directly from a microscopic theory.

\paragraph{Method} The most accurate and flexible microscopic approach to
superfluid dynamics in nuclear systems is density functional theory (DFT), 
which in principle is an exact approach.  Here we use an extension of 
Kohn-Sham DFT known as the time-dependent superfluid
local density approximation (TDSLDA), an orbital-based fermionic DFT that has been
proven to be very accurate for describing the dynamics of strongly correlated
fermionic systems in both ultracold atomic gases~\cite{PRL__2009,
  Science__2011, LNP__2012, PRL__2012, ARNPS__2013, PRL__2014, PRA__2015} and
in nuclear systems~\cite{PRC__2011, PRL__2015, PRL__2016, Mag2016}.  
In this approach, densities and the superfluid order parameter $\Delta$
are constructed from quasiparticle orbitals which are represented on a
3D lattice (without any symmetry restrictions) of size
$75\fm\times75\fm\times60\fm$ with lattice spacing corresponding to quite a 
large momentum cutoff $p_c\approx 400$ MeV/$c$,
and a volume that is sufficient to fit a single nucleus and a quantum vortex with
reasonable separation between the two.  To prevent vortices from neighboring
cells from interacting (due to the periodic boundary conditions), we introduce
a flat-bottomed external potential confining the system in a tube of a radius
$30\fm$. (See~\cite{Supplemental} and \cite{COCG} for details.) For initial states in our 
time-dependent simulations we chose stationary
self-consistent solutions of the TDSLDA with two constraints: i) the center of
mass of the protons is fixed at a specified position, ii) the phase of the
neutron pairing potential increases by $2\pi$
when moving around the center of the tube, i.e\@.
$\Delta(\rho, z, \phi) = \abs{\Delta(\rho,z)}\exp(\I\phi)$,
where $\rho=\sqrt{x^2+y^2}$
is the distance from the center of the tube and $\phi = \tan^{-1}\tfrac{y}{x}$.
We produce initial states for two background neutron densities,
$n=0.014\fm^{-3}$
and $n=0.031\fm^{-3}$,
with proton number $Z=50$.
These represent the zones 3 and 4 expected in neutron star crusts according to
the classification of Negele and Vautherin~\cite{NegeleVautherin}.  
Previous calculations are in clear disagreement in this region of densities.
We start the simulations from two configurations: an
\emph{unpinned} configuration where the nucleus is located outside the vortex,
close to the tube boundary, and a \emph{pinned} configuration where the nucleus
is located inside the vortex (see~\cite{Supplemental} for figures of these states).

The physics contained in the DFT is defined by the energy density functional
$\mathcal{E}$
which is a functional of the single-particle orbitals.  
For the
normal part we use the FaNDF\textsuperscript{0} functional constructed by
Fayans et al.~\cite{Fayans1, Fayans2}. It reproduces the infinite matter
equation of state of Refs.~\cite{FriedmanPandharipande, Wiringa}, many
properties of nuclei~\cite{PRL__2003a, arxiv:1507}, and allows one to construct
a very efficient solver of the TDSLDA equations (see~\cite{Supplemental}). The
only simplification we make is to omit the spin-orbit coupling term from the
functional as this greatly reduces the computational cost.  While the
spin-orbit term is important for finite nuclei, in the present context it is
not expected to significantly impact the final results.  The spin-orbit term
does not affect uniform matter, thus in our case where the ``nuclei'' are
embedded in a uniform gas of neutrons, it would shift the single-particle levels 
in the``nucleus'' in such a way as not to influence the physics of the vortex-nucleus system, 
as shown in~\cite{Avogadro2}.  These hardly influence the physics of
the vortex-nucleus system.  Likewise, since the depletion of the normal density
in the vortex core is small~\cite{PRL__2003}, the vortex density is approximately
uniform and one expects the influence of the spin-orbit term on the structure
of the vortex to be small. To the FaNDF\textsuperscript{0} functional we 
add a contribution describing the pairing correlations,
%
 $ \mathcal{E}_{\text{pair}}(\vect{r}) = g(n(\vect{r}))
    \abs{\nu_n(\vect{r})}^2 + g(p(\vect{r}))
    \abs{\nu_p(\vect{r})}^2 
  ,$
%
where $\nu_{p,n}$
are the $S = 0$
proton and neutron anomalous densities (proportional to the superfluid order
parameter and pairing gaps $\Delta_{n,p}$), $g$
is a density dependent coupling constant, and $n/p$ is density of
neutrons/protons. The coupling
constant $g$
is chosen so as to reproduce the neutron pairing gap in pure neutron
matter. It has the density dependence as predicted by BCS, but with maximum paring gap of $2\MeV$
(the full form is shown in~\cite{Supplemental}).  The local portion of the
anomalous densities $\nu_{n,p}$
diverges and requires regularization.  We use the procedure described in
Refs~\cite{PRL__2002, PRC__2002}, the accuracy of which has been validated
against a wide range of experimental results for cold atoms~\cite{ARNPS__2013,
  PRL__2003b, PRA__2007, PRL__2008, PRL__2009, Science__2011, PRL__2012,
  PRL__2014, PRA__2015} and nuclear problems~\cite{PRL__2016, PRC__2011,
  PRL__2003, PRL__2003a}. 

The TDSLDA approach automatically includes various dissipative processes,
including superfluid and normal phonon excitations, Cooper pair breaking, and
Landau damping.  These are crucial for a correct description of vortex pinning
and unpinning \cite{Sedrakian}.  Consider pinning: for a nucleus to capture a
vortex, the vortex must dissipate its collective energy, otherwise it will
simply orbit the nucleus as governed by the Magnus force, like a precessing
spinning top.  We demonstrated in~\cite{PRA__2015} that the TDSLDA accurately
models the formation and decay of solitonic defects -- from domain walls into
vortex rings and vortex lines.  These effects cannot be reproduced without
dissipation, and the agreement with experiments~\cite{MIT1, MIT2} validates
that the so-called one-body dissipation naturally present in the TDSLDA is
sufficient to correctly capture vortex dynamics.  With the TDSLDA approach, we
can thus extract both the magnitude of the vortex-nucleus interaction as well
as the dynamical time scales.

\begin{figure}[t]
\includegraphics[width=1.0\columnwidth]{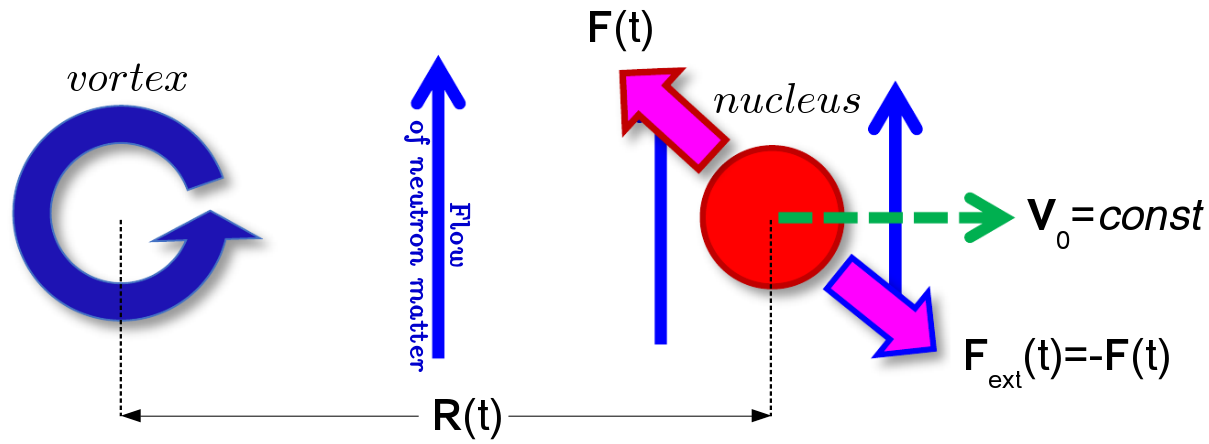}
\caption{(Color online) Schematic figure explaining the method used for the
  force extraction. The force $\vect{F}$ depends on relative distance between
  vortex and nucleus $R$, moving with a constant velocity $\vect{v}_0$. The external 
  force $\vect{F}_{\textrm{ext}}(t)$ is chosen to compensate exactly $\vect{F}$.
  \label{fig:method_idea}}
\end{figure}
\begin{figure*}[th]
\newcommand\includeframe[1]{%
  \includegraphics[width=0.5\columnwidth, trim=140 75 110 103, clip]{#1}}
\includeframe{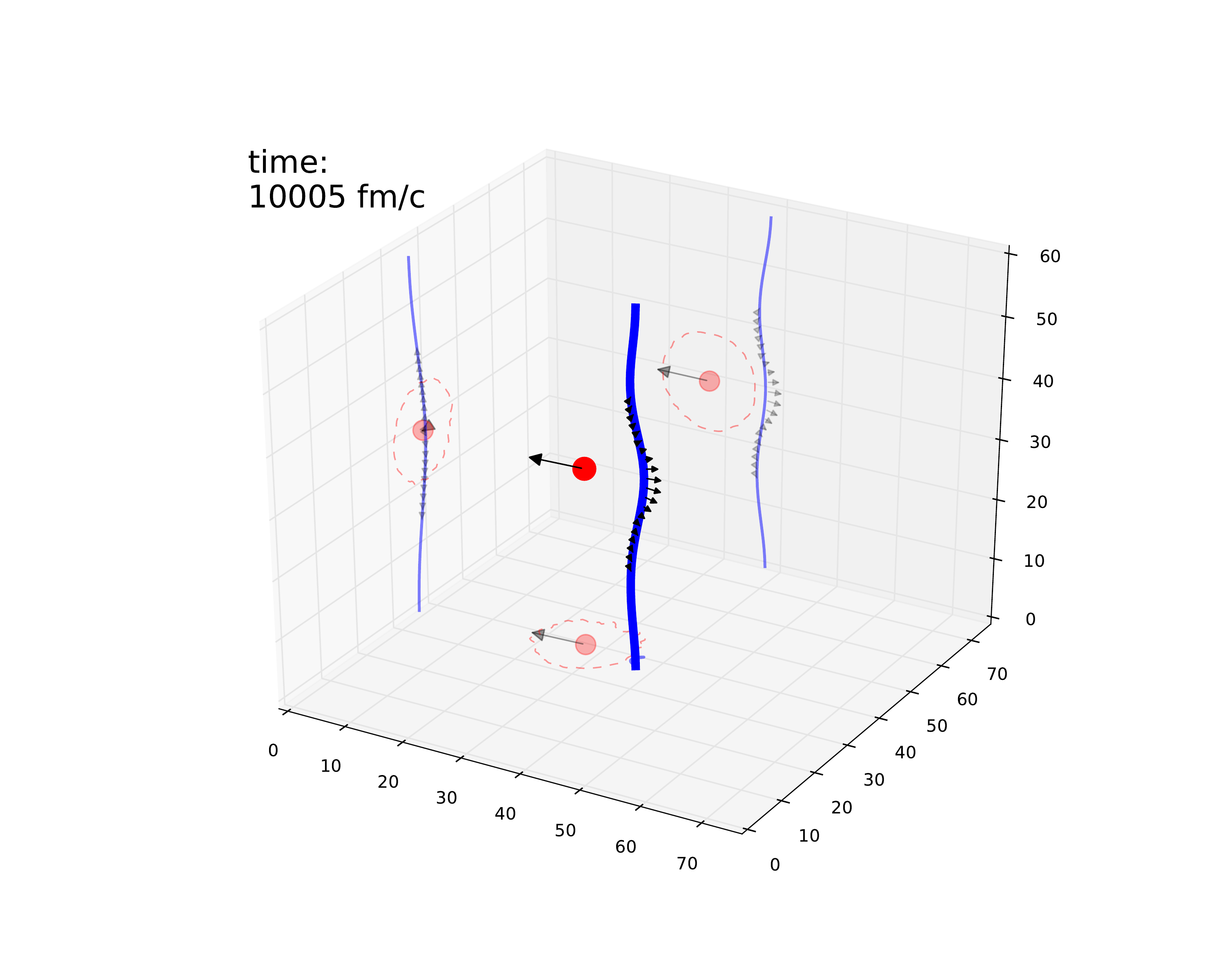}%
\includeframe{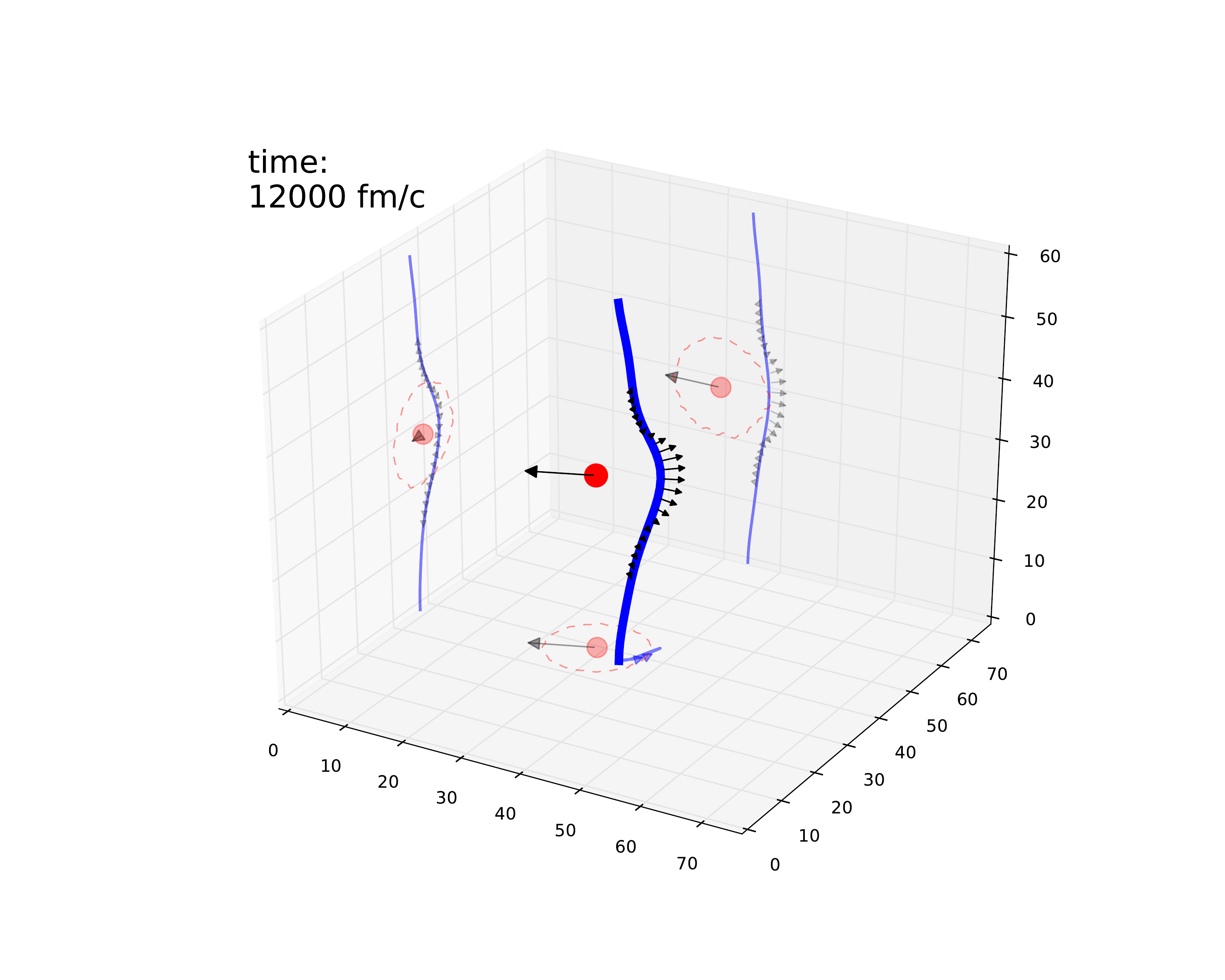}%
\includeframe{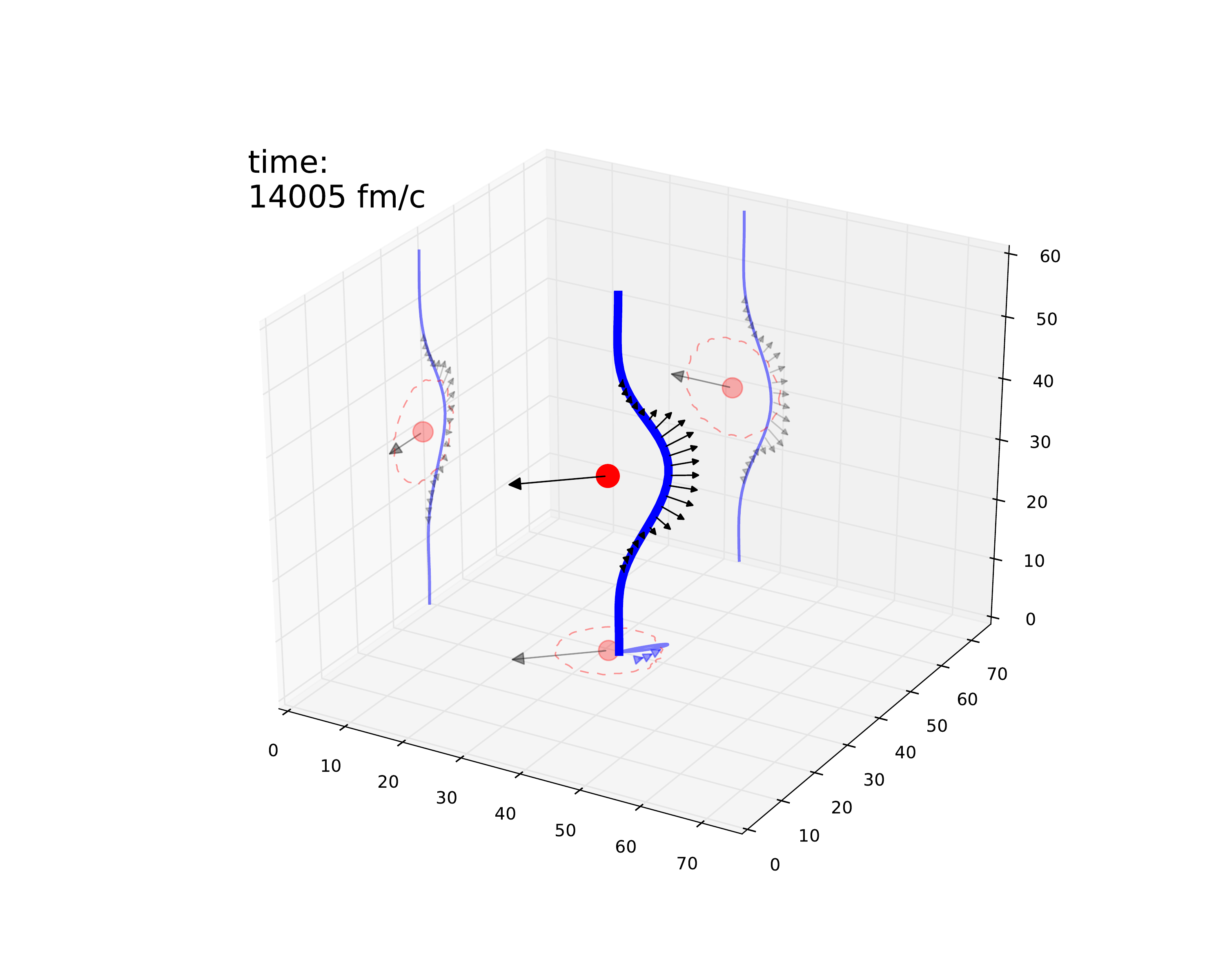}%
\includeframe{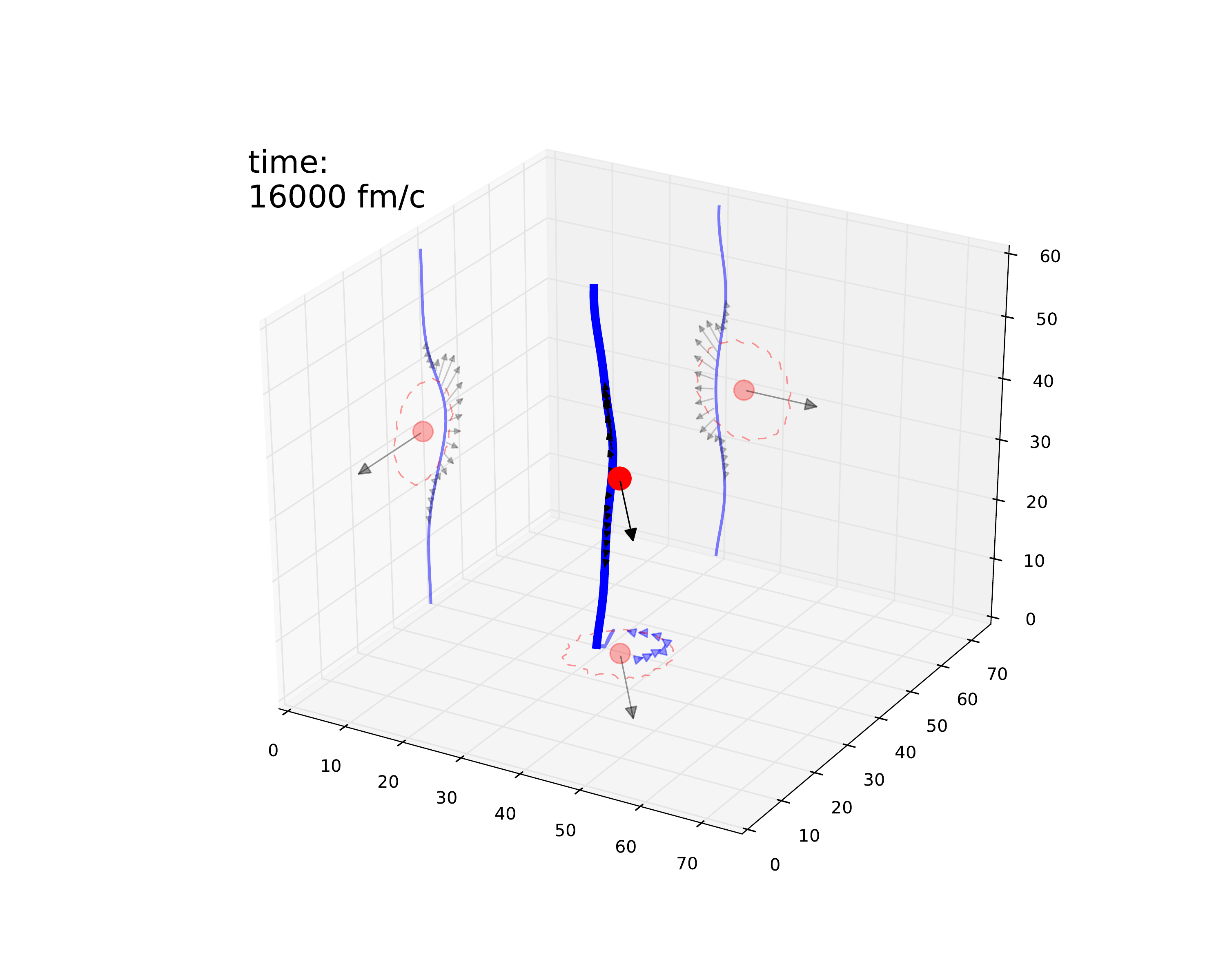}\\
\includeframe{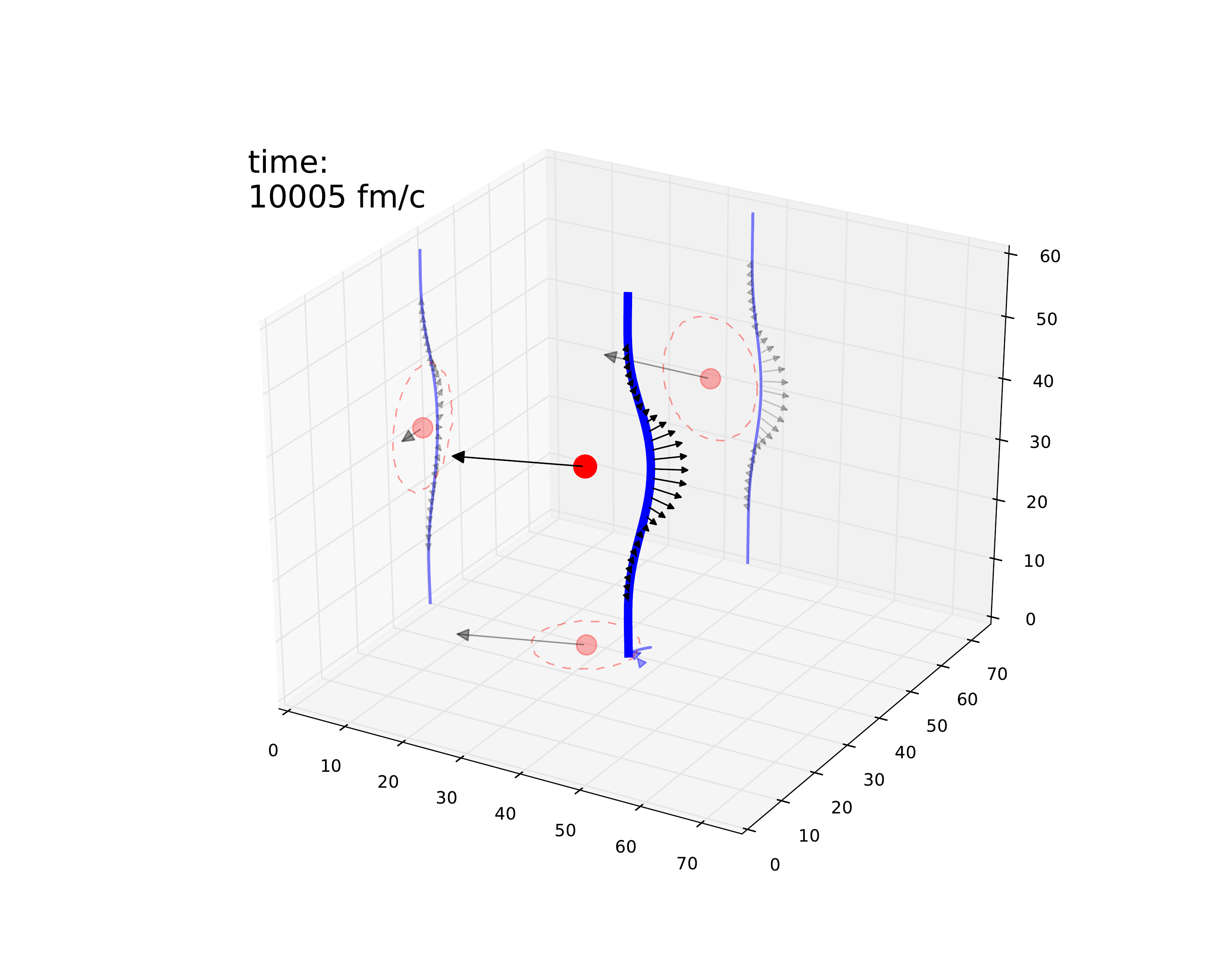}%
\includeframe{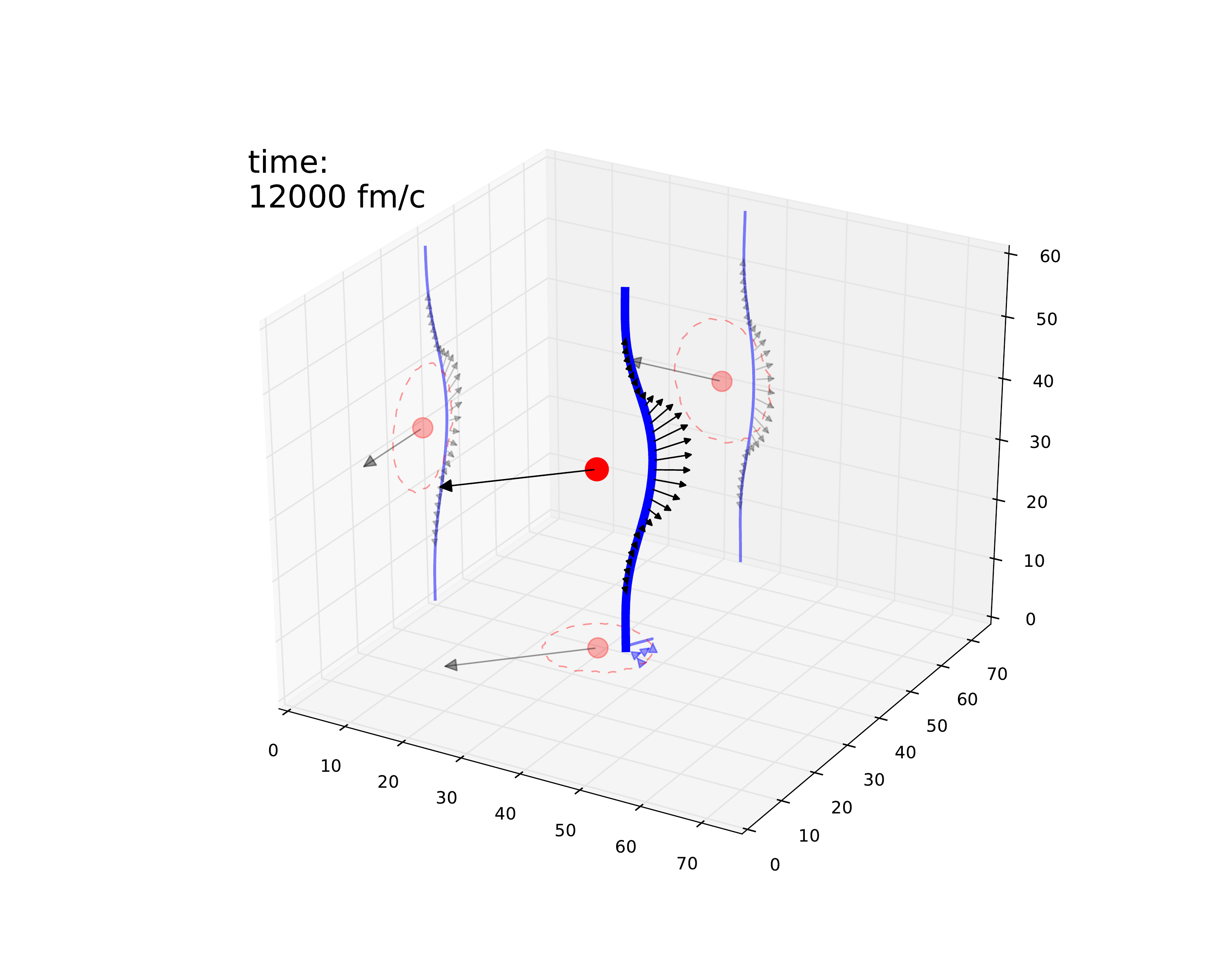}%
\includeframe{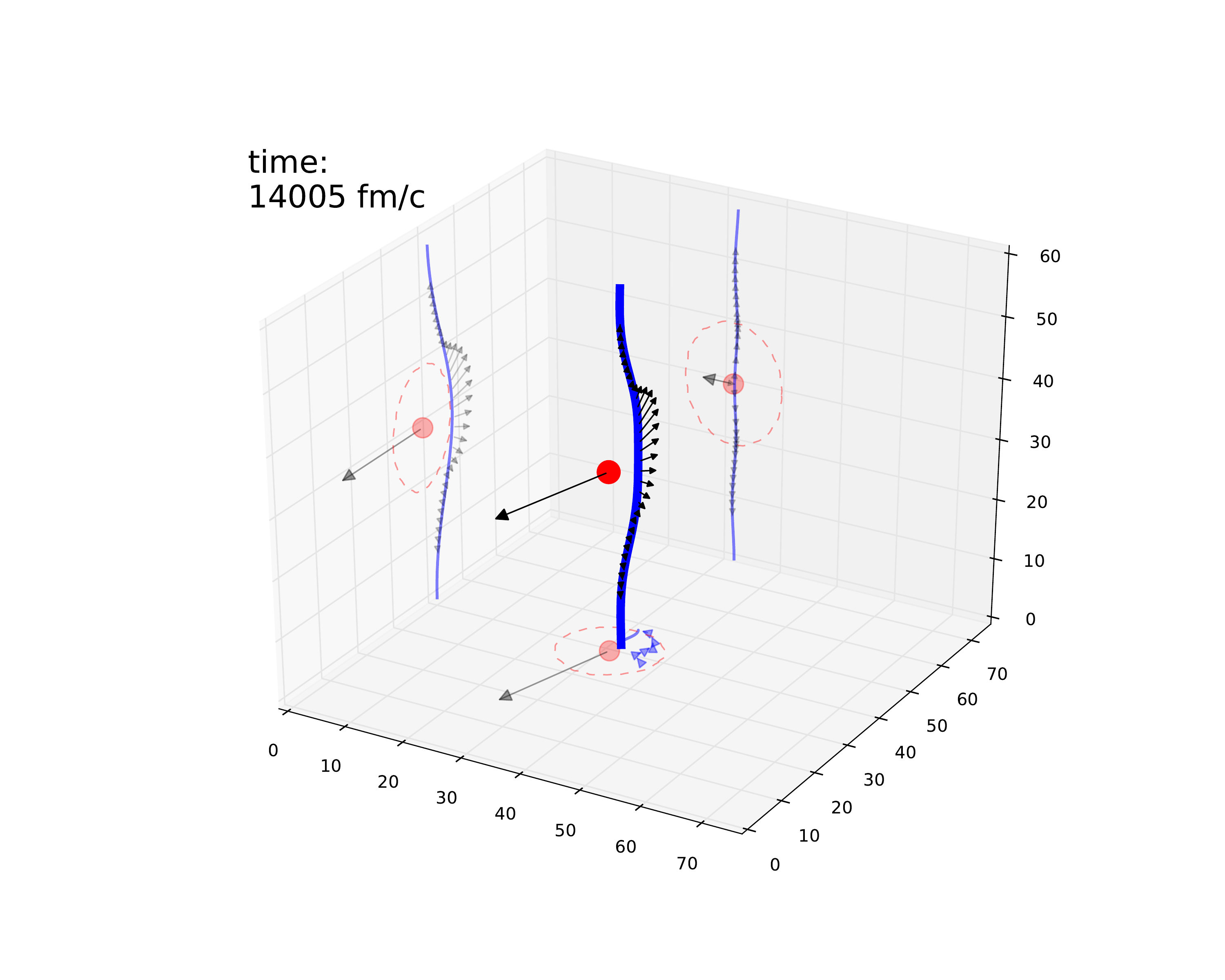}%
\includeframe{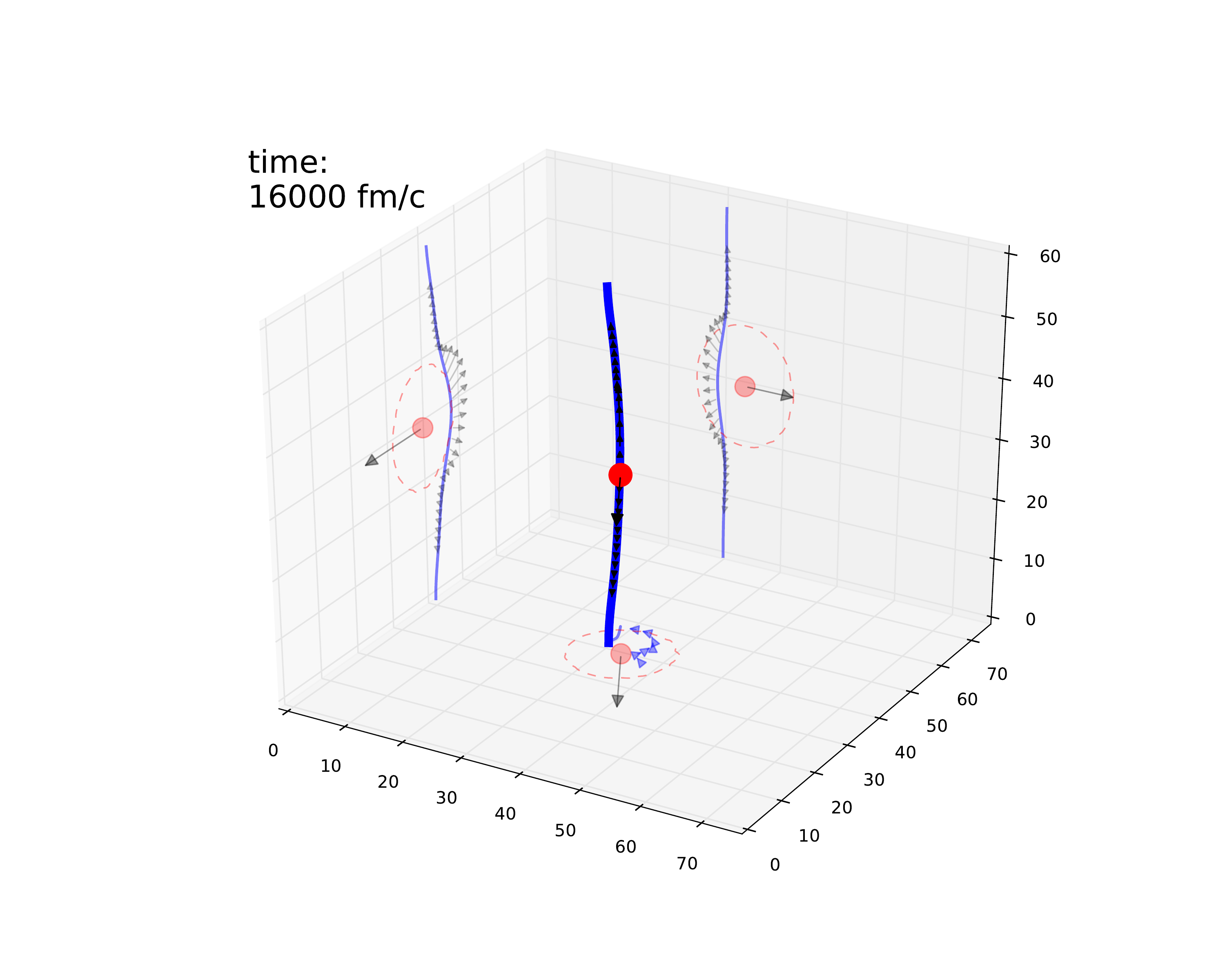}
\let\includeframe\undefined
\caption{(Color online) Dynamics of the system for times corresponding to small
  vortex-nucleus separations for neutron matter density $n=0.014\fm^{-3}$ (top)
  and $0.031\fm^{-3}$ (bottom). Frames from left to right correspond to times 
  $(10,12,14,16)\times 1,000\fm/c$ (for full movies see~\cite{Supplemental}). 
  Blue line indicates
  the vortex core position extracted from the order parameter $\Delta$
  (see~\cite{Supplemental} for details). Red dot indicates position of the
  center of mass of protons. The vector attached to the red dot denotes the
  vortex-nucleus force $\vect{F}(R)$. Vectors attached to the vortex indicate   contributions to the force $-d\vect{F}$ extracted from force per unit length, see Eq.~(\ref{eqn:f_r}) and inset (a) of Fig.~\ref{fig:force_per_unit_length}. They are scaled by factor 3 for better visibility. Projections of the view are shown on sides
  of the box. Red dashed lines denote shape of nucleus (defined as a point
  where density of protons drops to value $0.005\fm^{-3}$). By blue triangles 
  (on XY-plane) trajectory of the vortex up to given time is shown. 
  \label{fig:movieframes}}
\end{figure*}

To extract the effective force between a quantized vortex and a nucleus, we
apply Newton's laws.  Suppose that only two forces act on the nucleus: the
force $\vect{F}$
arising from the interaction with the vortex and a known external force
$\vect{F}_{\text{ext}}$.
In the simplest case, the vortex-nucleus force depends on the relative distance
between interacting objects $R$.
If the nucleus moves with a constant velocity $\vect{v}_{0}$
which is below the critical velocity (so that phonons are not excited), then
the relation $\vect{F}(t)=-\vect{F}_{\text{ext}}(t)$
holds. Combining this information with the relative distance $R(t)$
we can extract the vortex-nucleus force as a function of the separation
$\vect{F}(R)$,
see Fig.~\ref{fig:method_idea}. We choose the external force to be constant in
space and acting only on the protons.  This force moves the center of mass of
the protons together with those neutrons bound (entrained) in the nucleus
without significantly modifying the internal structure of the nucleus or
surrounding neutron medium.  We adjust the force to ensure that the center of
mass of the protons moves with constant velocity $\vect{v}_{0}$:
\begin{equation}
 \vect{F}_{\textrm{ext}}(t+\Delta t) = \vect{F}_{\textrm{ext}}(t) - \alpha\left[
   \vect{v}(t) - \vect{v}_{0}\right],
\end{equation} 
where $\vect{v}(t)$
is the velocity of the center of mass of protons and $\alpha$
is the coefficient governing the rate of adjusting the force. In our
simulations we dragged the nucleus with a very small velocity
$v_0=0.001c$
along the $x$
axis to ensure that no phonons are excited. The velocity is far below the
critical velocity of the system and is sufficiently small that the systems
follow an almost adiabatic path.

\paragraph{Results of dynamical simulations} In the first set of simulations,
we start from an unpinned configuration and drag the nucleus towards the
vortex.  Fig.~\ref{fig:movieframes} shows the time evolution of these systems
for small vortex-nucleus separations (see~\cite{Supplemental} for movies of the
entire simulations). As the nucleus approaches to the vortex, it exerts a force
$\vect{F}(R)$
on the vortex which responds by moving according to the Magnus relationship
$\vect{F}_M \propto \vect{\kappa} \times \frac{\partial\vect{r}}{\partial t}$,
where $\vect{r}$
specifies the vortex-core position.  The vortex is initially moving
perpendicular to this force along positive $y$ direction
visually confirming that the force is indeed repulsive and initially directed
along the $x$
axis away from the nucleus. In the case of attraction the vortex would
initially move along the negative $y$
direction.  For both densities considered, the vortex-nucleus interaction is
clearly repulsive and increasing with density, a result in agreement with the
hydrodynamic approximation~\cite{Supplemental}.  The curvature of the vortex
bending for the closest vortex-nucleus configuration is set by the nucleus.
(There is also a small displacement of both ends of the vortex during the
evolution in our simulation box.) For the higher density, the vortex induces
visible nuclear prolate deformation with the elongation axis set by the vortex
axis.
To confirm the repulsive nature of the force at very small vortex-nucleus
separations, we also start simulations from ``pinned'' configurations. In both
cases the vortex rapidly unpins (with a timescale shorter than $1,000\fm/c$), 
i.e\@., the vortex is immediately expelled from the nucleus, indicating that
the pinned configuration is dynamically unstable.  The initial energy is
transferred into stretching of the vortex line as it bows out away from the
nucleus.

We will proceed now to estimate the vortex tension $T$.
The vortex is the longest at $t_{\textrm{max}}\approx 14,000\fm/c$
for both low and high densities, $0.014\fm^{-3}$
and $0.031\fm^{-3}$
respectively.  The length of the vortex increases by
$\Delta L=L(t_{\textrm{max}})-L(0) = 3.5\fm$ and
$1.5\fm$ and the total excitation energy of the system is
$E^*=E(t_{\textrm{max}})-E(0) = 5$~MeV and
$11$~MeV respectively.
Assuming all of this energy is stored in the vortex, we obtain an upper bound
on the vortex tension of $T\lesssim 1.4\;\textrm{MeV/fm}$
 and $7.3\;\textrm{MeV/fm}$, respectively.
The energy of a vortex line in the leading order hydrodynamic approximation is
$E \approx \rho_s \kappa^2 L \ln (D/2\xi)/4\pi$~\cite{Supplemental}, 
where $D$ is the diameter
of the simulation cell, which has to be replaced with 
the average vortex separation $l_v$ in the neutron star crust~\cite{Andersson}.  This simple
hydrodynamic approximation suggests that different tensions arise from changes in the 
neutron superfluid density $\rho_s$ and vortex core size $\xi$.
Estimating $\rho_s \sim n$ gives a ratio of $0.77$~\cite{Seveso},
which is much larger than the ratio $1.4/7.3\approx 0.18$
obtained from our microscopic simulations.  At higher densities the 
vortex is thus much stiffer then expected from
hydrodynamic estimates. 

\begin{figure}[t]
\includegraphics[width=1.0\columnwidth]{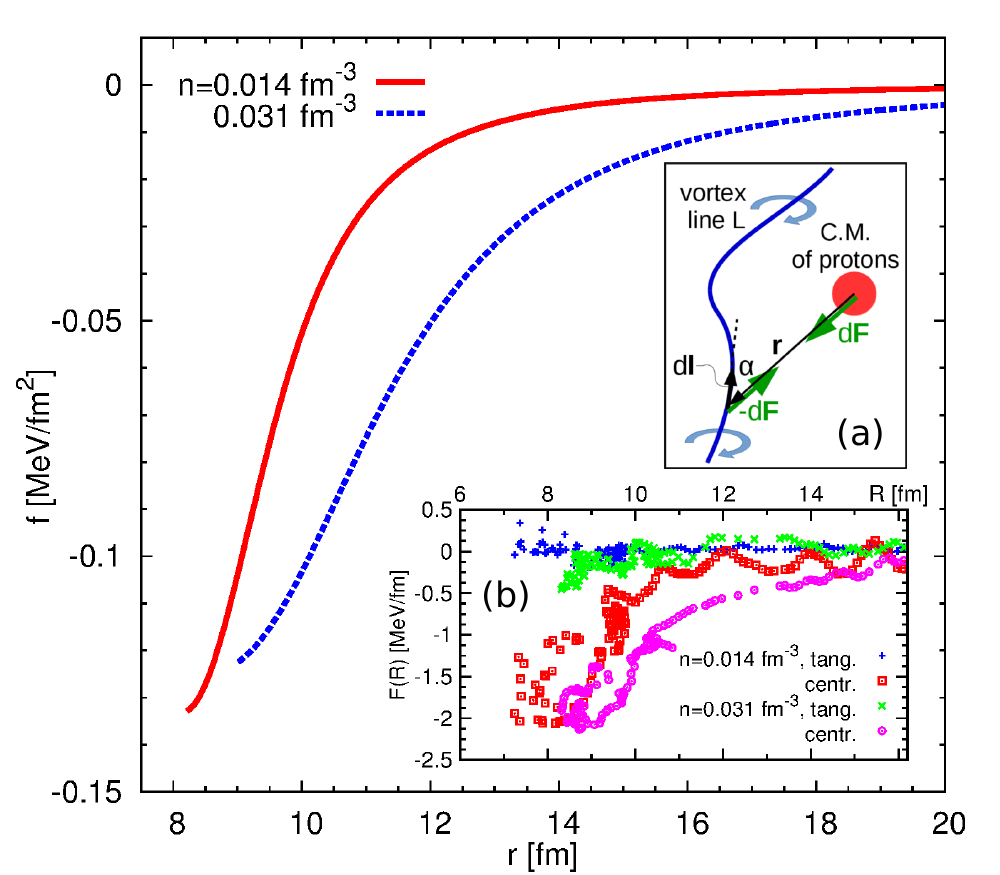}
\caption{ (Color online) Extracted force per unit length $f(r)$
  for both densities. Negative values correspond to the repulsive force. In
  inset (a) the schematic configuration is shown explaining the extraction
  procedure according to Eq.~(\ref{eqn:f_r}). Inset (b) shows the measured
  total force $F(R)$
  as shown in Fig.~\ref{fig:method_idea} for both densities. The force has been
  decomposed into tangential and centripetal components with respect to the
  instantaneous vortex position.
\label{fig:force_per_unit_length}}
\end{figure}

\paragraph{Force per unit vortex length} 
Combining the information about the force $\vect{F}(t)$
with the vortex-nucleus separation $R(t)$,
we extract the force for various separations $R$,
defined as the distance within the plane perpendicular to the symmetry axis of
the confining tube.  We decompose the force into a tangential and a centripetal
components with respect to the vortex position at each time.  These results are
presented in the inset (b) of Fig.~\ref{fig:force_per_unit_length}.
The extracted force is predominantly central
with a negligible tangential component.
The effective range of the force is about $10\fm$
for the lower density, increasing to about $15\fm$
for the higher density, consistent with an increasing coherence length $\xi$ 
with density and decreasing neutron pairing gap. 
The behavior of the total force for small separations
demonstrates that it is not merely a function of a distance.  At small
separations, the deformation of the vortex line and the nuclear deformation 
become important degrees of freedom.

To characterize the effects of the vortex geometry, we extract the force per
unit length $f(r)$.
Inspired by the vortex filament model (see~\cite{VinenNiemela, Tsubota} and
references therein), we divide the vortex line into elements of length
$\d\vect{l}$.  Each element exerts force on a nucleus
\begin{equation}
  \d\vect{F} = f(r)\sin\alpha \uvect{r} \d{l},\label{eqn:f_r}
\end{equation} 
where $\vect{r}$
denotes the position of the vortex line element from the center of mass of the
protons, $\alpha$
is angle between vectors $\d\vect{l}$
and $\vect{r}$
(see inset (a) of Fig.~\ref{fig:force_per_unit_length}), and
$\uvect{r} = \vect{r}/r$.
The force $\vect{f}_{\text{VN}}$ in Eq.~(\ref{eqn:vmotion}) is given by $-f(r)\sin\alpha \uvect{r}$.
The total force is the sum of contributions from all vortex elements
$\vect{F}=\int_{L}\d\vect{F}$.
We model $f(r)$
with a Pad\'{e} approximant with the asymptotic behavior
$f(r\rightarrow\infty)\propto r^{-3}$
consistent with hydrodynamic predictions.  The parameters of the Pad\'{e}
approximant are determined from a least-squares fit to all simulation data
resulting in the force per unit length $f(r)$
characterization of the vortex-nucleus interaction shown in
Fig.~\ref{fig:force_per_unit_length}.  (The hydrodynamic results and fitting
procedure are described in detail in~\cite{Supplemental}.)  \exclude{ To
  parametrize $f(r)$ we have used Pad\'{e} approximant
\begin{equation}
 f(r) = \dfrac{\sum_{k=0}^{n}a_k r^k}{1+\sum_{k=1}^{n+3}b_k r^k},\label{eqn:f_r_Pade}
\end{equation} 
where we imposed asymptotic behavior $f(r\rightarrow\infty)\propto 1/r^{3}$,
predicted by irrotational, incompressible hydrodynamics
(see~\cite{Supplemental}).  We have constructed the least square problem and
minimized residuals defined as a difference of measured force and the
prediction~(\ref{eqn:f_r}) with respect to parameters $\{a_k\}$
and $\{b_k\}$.
We have found that it is sufficient to use Pad\'{e}
approximant~(\ref{eqn:f_r_Pade}) with $n=2$
(see~\cite{Supplemental} for convergence tests). Results of reconstruction are
shown in
} 
This simple characterization of the force $\vect{F}$
works better at lower densities, which is consistent with the
larger nuclear deformations seen at higher densities. Nuclear deformations 
introduce an orientation dependence to the force that is not captured by the simple
model~\eqref{eqn:f_r}.

\paragraph{Conclusions} We have performed unconstrained simulations of a
quantum vortex dynamics in superfluid neutron medium in the presence of a
nucleus using an appropriate time-dependent extension of DFT to superfluid
system. We have determined that the vortex-nucleus force is repulsive and
increasing in magnitude with density for the densities characteristic of the
neutron star crust ($0.014$
and $0.031\fm^{-3}$).
The vortex line shape is strongly affected by the interaction at small
separations, leading to significant bending and its lengthening, controlled by
the size of the nucleus.  Moreover, the vortex-nucleus interaction also induces a
deformation of the nucleus.  These results demonstrate that the
vortex-nucleus interaction cannot be described by a function of their
separation alone. To fully characterize 
the vortex-nucleus interaction we have extracted the force per
unit length for various vortex-nucleus configurations.  
For  velocities of any ambient flow in the background smaller than 
$v_c\sim (1-5)\times10^{-4}c$
the pinned superfluid can store enough angular momentum to
drive the giant glitches seen in pulsars~\cite{BLink}.

\begin{acknowledgments}
  We are grateful to K.~J.~Roche and A.~Sedrakian for helpful discussions and
  comments.  We thank Witold Rudnicki,
  Franciszek Rakowski, Maciej Marchwiany and Kajetan Dutka from the
  Interdisciplinary Centre for Mathematical and Computational Modelling (ICM)
  of Warsaw University for useful discussions concerning the code optimization.
  This work was supported by the
  Polish National Science Center (NCN) under Contracts
  No. UMO-2013/08/A/ST3/00708. The method used for generating initial states was
  developed under a grant of Polish NCN under Contracts
  No. UMO-2014/13/D/ST3/01940.  Calculations have been performed: at the OLCF
  Titan - resources of the Oak Ridge Leadership Computing Facility, which is a
  DOE Office of Science User Facility supported under Contract
  DE-AC05-00OR22725; at NERSC Edison - resources of the National Energy
  Research Scientific computing Center, which is supported by the Office of
  Science of the U.S. Department of Energy under Contract
  No. DE-AC02-05CH11231; at HA-PACS (PACS-VIII) system - resources provided by
  Interdisciplinary Computational Science Program in Center for Computational
  Sciences, University of Tsukuba. This work was supported in part by U.S. DOE
  Office of Science Grant DE-FG02-97ER41014, and a WSU New Faculty Seed Grant.
\end{acknowledgments}


\newpage
\begin{center}
{\bf Supplemental online material for:}\\
{\bf ``Vortex pinning and dynamics in the neutron star crust''}\\
\end{center}
\begin{small}
\noindent
In the supplemental material various technical aspects are discussed related to the generation of initial configurations, integration of TDSLDA equations, and the vortex detection algorithm. 
We also provide details concerning the fitting procedure, which allows to extract the force per unit length acting on the vortex,
and the derivation of the asymptotic expression of the vortex-impurity interaction from irrotational and incompressible hydrodynamics. 
\end{small}

\section{Initial configurations}

Initial states are prepared as self-consistent solutions of SLDA equations
\begin{equation}
  \begin{pmatrix}
    h  & \Delta \\
    \Delta^* & -h
  \end{pmatrix}
  \begin{pmatrix}
    u_k \\
    v_k
  \end{pmatrix}=
  \varepsilon_k
  \left( 
    \begin{array}{l}
      u_k \\ 
      v_k            
    \end{array}
  \right) ,
  \label{eqn:staticSLDA}
\end{equation} 
where the single-particle Hamiltonian $h$
and pairing potential $\Delta$
are obtained by taking the appropriate functional derivatives of the energy
density functional, and $\varepsilon_k$ are single-quasiparticle energies.
Standard self-consistent approach requires a series of diagonalizations of the single-quasiparticle Hamiltonian.
For the case without the spin-orbit interaction, the Hamiltonian is represented by a matrix of size $(2\times 50\times 50\times 40)^2=200,000^2$ (factor $2$ corresponds to $u$ and $v$ components of the wave function). Single diagonalization of this matrix (for protons and neutrons) takes about $40$ min. on Edison supercomputer (NERSC)~\cite{EdisonSM} with 36,864 cores. Taking into account that more than $100$ iterations are required to get 
a reasonably convergent solution, it gives an unacceptable computational cost of the order of few millions of CPU hours per initial state. For this reason we have used alternative methods to decrease the number of diagonalizations. 
Namely, we started the process from the uniform matter solution using the real time dynamics (with a desired number of protons and neutrons), and subsequently we slowly turned on the confining potential for neutrons 
and the harmonic potential for protons. 
The harmonic potential localizes protons in a desired position in space. The real-time dynamics is supplemented by potential simulating quantum friction (see~\cite{QuantumFrictionSM} for details). During this stage quite accurate initial profiles of densities defining the single-quasiparticle Hamiltonian have been created, for a state without vortex. Next, we started self-consistent iterations to compute densities from Green's functions, using the method very similar to the one applied in studies of large electronic systems~\cite{RTakayamaSM,SYamamotoSM}. The method extracts densities without explicit diagonalization of HFB matrix (for details see~\cite{COCGSupp}). To get a state with a vortex in each iteration we imprinted the correct phase dependence on 
the neutron paring potential $\Delta$. Once the convergence (with a satisfactory accuracy) has been reached we performed a diagonalization in order to extract wave functions. Finally, we performed short real-time dynamics with the quantum friction potential in order to get rid of residual excitations. This methodology allows us to generate an initial states invoking the diagonalization routine only once. 

The neutrons are confined in axially symmetric external potential ($U_{\textrm{ext}}$ in MeV, $r$ in fm)
\begin{equation}
 U_{\textrm{ext}}(r)=\left\lbrace 
 \begin{array}{ll}
  0, & r\leqslant 30\\
  50\,s(r-30,5,2), & 30<r<35\\
  50, & r\geqslant 35
 \end{array}
\right. 
\end{equation} 
where $s$ denotes the switching function,
\begin{equation}
 s(x,w,\alpha)=\dfrac{1}{2}+\dfrac{1}{2}\tanh\left[ \alpha\tan\left(  \frac{\pi x}{w}-\frac{\pi}{2} \right) \right].
\end{equation} 
$r=\sqrt{x^2+y^2}$ is the distance from the symmetry $z$-axis.
The potential allows to avoid the interference of the quantum vortex with other vortices from neighboring cells.
Note that inside the tube the potential is perfectly flat and it does not affect vortex dynamics.  Dynamics of a nucleus is also not affected, as long as the nucleus does not touch the boundary. 

\begin{figure}[t]
\includegraphics[width=1.0\columnwidth]{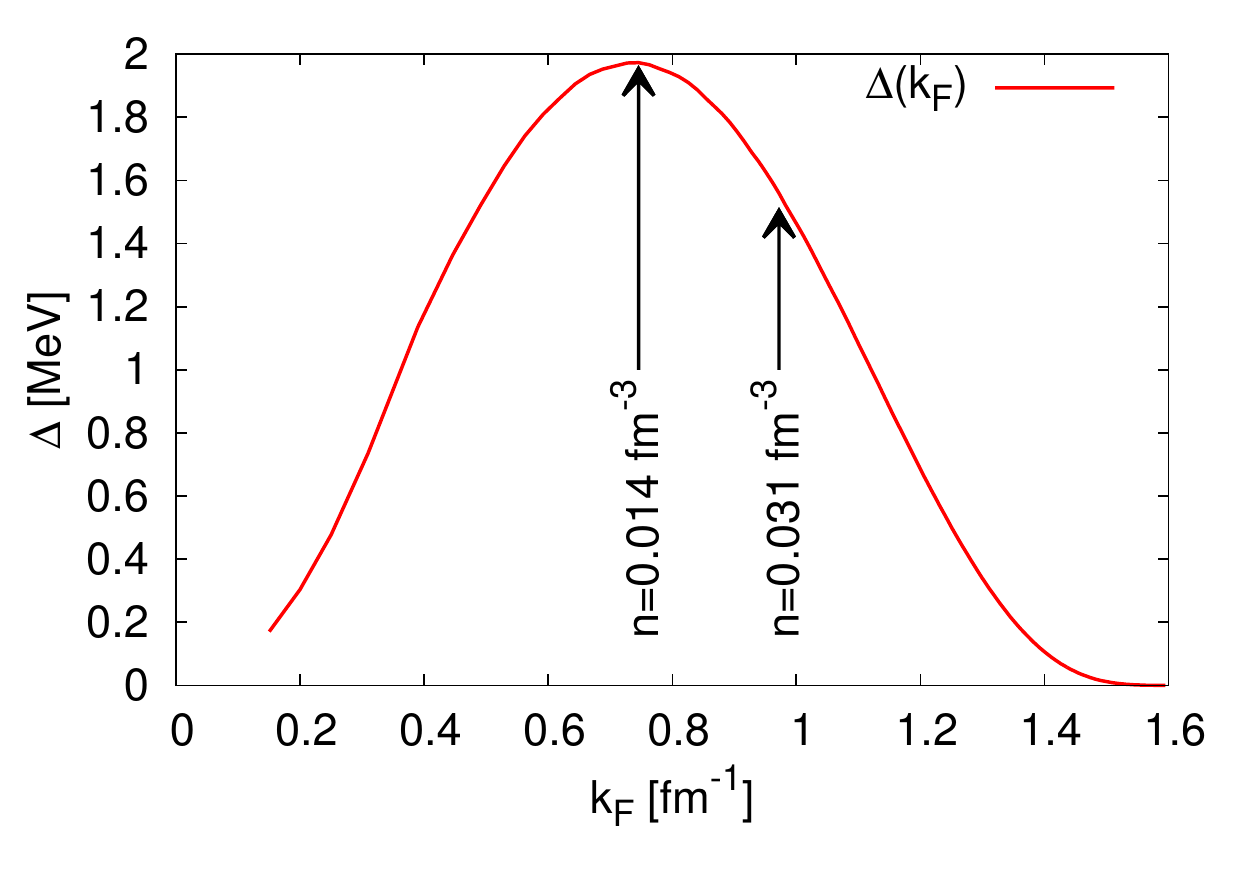}
\caption{ (Color online) The neutron pairing gap as a function of the Fermi momentum $\kF=(3\pi^2 n)^{1/3}$ used in calculations. Arrows indicate points for which simulations have been performed: 
$\kF=0.75\fm^{-1}$ and $0.97\fm^{-1}$, respectively.  
\label{fig:delta}}
\end{figure}

In order to fit the pairing coupling constant $g$, we have solved static SLDA equations~(\ref{eqn:staticSLDA}) for uniform neutron matter for various densities. For each density we treated $g$ as a free parameter and we adjusted it 
to reproduce the value of the neutron pairing gap. 
The standard BCS theory of superconductivity predicts the density
dependence of the paring gap $\Delta$ in pure neutron matter,
but overestimates its amplitude (see, e.g., Ref.~\cite{GandolfiSM} and references
therein). We therefore scale the BCS results to obtain maximum paring gap of
$2\MeV$. Note that in our approach both
neutrons and protons are superfluid, and that we assume isospin is a good
symmetry for the pairing channel (both couplings are equal).
In Fig.~\ref{fig:delta} we present the neutron pairing gap as a function of the Fermi momentum used in the fitting procedure. After tabulating coupling constants $g$ as a function of density $n$ we constructed an interpolating function which has been used subsequently in dynamical simulations.

\begin{figure}[ht]
\includegraphics[width=0.8\columnwidth]{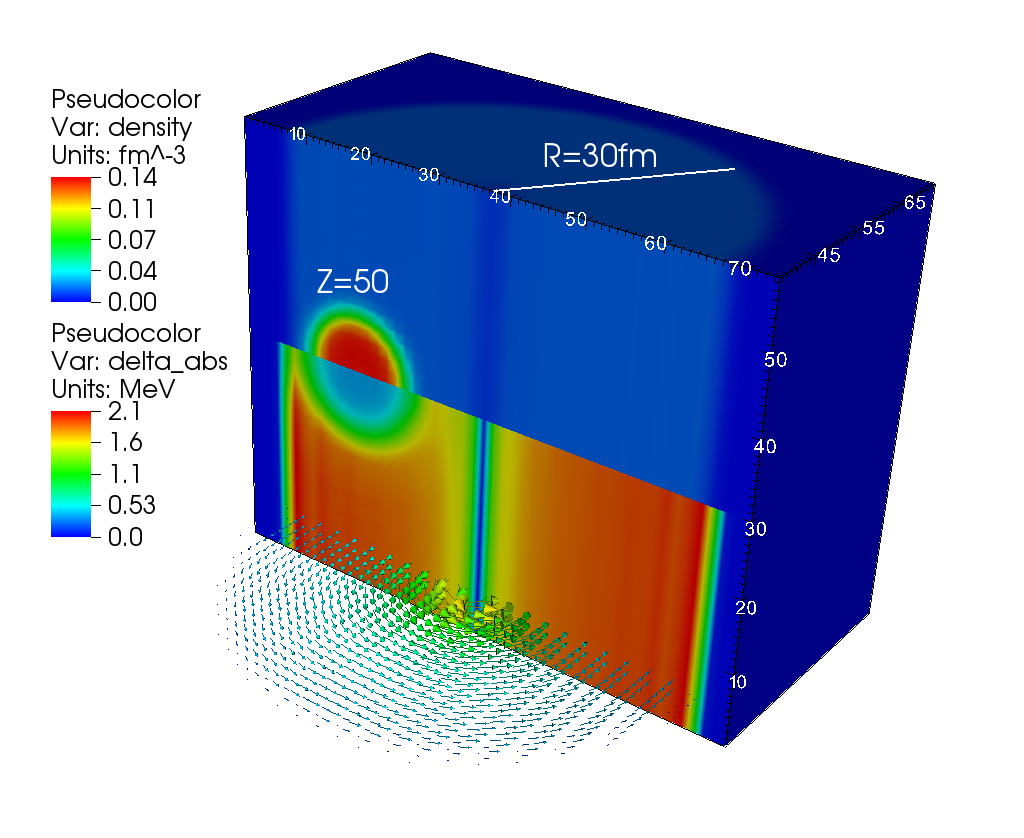}
\caption{ (Color online) Example of initial unpinned configuration for $n=0.014\fm^{-3}$. The upper part of the box shows the total density distributions, while the lower part shows the absolute value of the neutron paring potential $\Delta$. The vanishing pairing field and the depletion of the density along the tube symmetry axis are due to the presence of a quantum vortex.  Arrows in the bottom part of the figure show the circular flow of neutrons. 
\label{fig:setup}}
\end{figure}

In Fig.~\ref{fig:setup} an example of the initial state used in the simulations is shown. Within the box we are able to fit a nucleus and a quantum vortex. It also provides enough space to study the interaction at variety of vortex-nucleus mutual arrangements.

\begin{figure}[th]
\includegraphics[width=1.0\columnwidth]{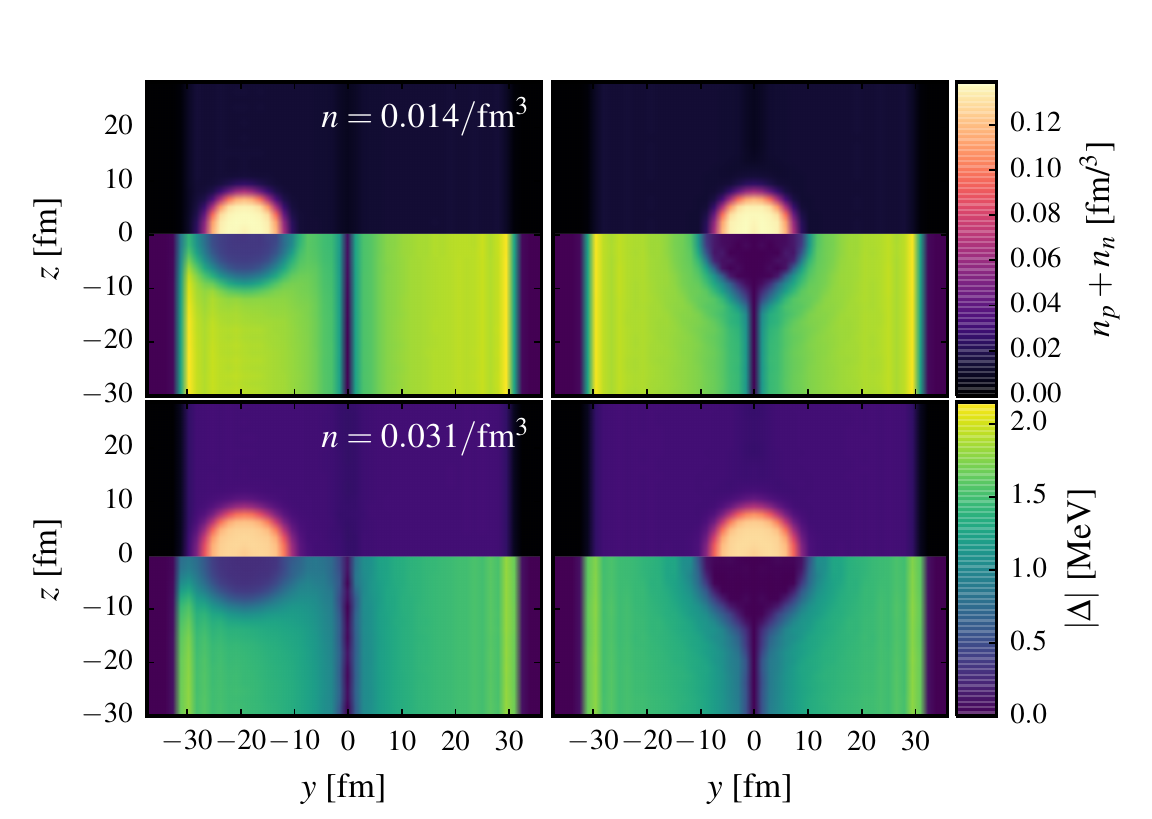}
\caption{ (Color online) Initial states used for dynamic simulations for neutron background density $n=0.014\fm^{-3}$ (top) and $0.031\fm^{-3}$ (bottom). In each box the total density distribution (upper half) and  
the absolute value of the neutron paring potential (lower half) are shown.
\label{fig:initstates_frames}}
\end{figure}

In order to examine the possibilities of both
pinning and anti-pinning scenario, we have generated initial states for the background neutron density
$n=0.014\fm^{-3}$
and $0.031\fm^{-3}$,
both for pinned and unpinned configurations.  Number of protons is set to be
$50$,
while number of neutrons is about $2,530$
and $5,710$,
respectively. For both densities the energy per particle turned out to be
larger for pinned configuration than for the unpinned one.  The differences in
energy per particle are about $6\;\textrm{keV}$
and $4\;\textrm{keV}$,
respectively for densities $n=0.014\fm^{-3}$
and $0.031\fm^{-3}$.
A simple estimate of the pinning energy computed as the energy per particle times the average number of
particles gives values of $15\MeV$
and $22\MeV$
respectively.  These numbers are close to values reported
in~\cite{Avogadro2Supp} for the SkM$*$
and SGII interactions which have an effective mass close that of the
$\textrm{FaNDF}^0$
functional we used.  We emphasize, however, that these are only estimates.  Our
initial states are constructed for time-evolution, not for computing energy
differences (the pinning energy), and have slightly different numbers of
particles which will introduce additional corrections to the quantitative
pinning energy.  In Fig.~\ref{fig:initstates_frames} we show the density
distribution and the neutron pairing potential for generated initial states.

\section{Integration of TDSLDA equations - technical aspects}

The TDSLDA equations are formally equivalent to the time-dependent
Hartree-Fock-Bogoliubov (TDHFB) in coordinate representation, or
time-dependent Bogolubov-de Gennes (TD BdG) equations
\begin{equation}
  i \hbar \dfrac{\partial}{\partial t}
  \left( 
    \begin{array}{l}
      u_k \\ 
      v_k            
    \end{array}
  \right) 
  =
  \begin{pmatrix}
    h  & \Delta \\
    \Delta^* & -h
  \end{pmatrix}
  \begin{pmatrix}
    u_k \\
    v_k
  \end{pmatrix}.
  \label{eqn:TDDFT}
\end{equation}
We have integrated these equations using the symplectic splitting operator method.
The time stepping algorithm consists of the following operations:
\begin{equation}
    \left( 
    \begin{array}{l}
      u_k(t+\Delta t) \\ 
      v_k(t+\Delta t)            
    \end{array}
    \right) =\exp\left(-i\mathcal{H}\Delta t/\hbar \right)     \left( 
    \begin{array}{l}
      u_k(t) \\ 
      v_k(t)            
    \end{array}
    \right),
\end{equation} 
where the matrix $\mathcal{H}(n(t),\nu(t),...)$ depends on wave-functions through densities. The stepping algorithm reaches the highest accuracy if $\mathcal{H}$ is provided for midpoint time $t+\frac{\Delta t}{2}$. We use Heun's method to  produce midpoint Hamiltonian, i.e. first we performed a trial step $\varphi_{\textrm{trial}}(t+\Delta t)=\exp\left[ -i\mathcal{H}(t)\Delta t/\hbar \right] \varphi(t)$, then from $\varphi_{\textrm{trial}}$ we computed densities and formed $\mathcal{H}_{\textrm{trial}}(t+\Delta t)$. Finally we approximated $\mathcal{H}(t+\frac{\Delta t}{2})\cong\frac{1}{2}\left[ \mathcal{H}(t)+\mathcal{H}_{\textrm{trial}}(t+\Delta t)\right]$. 
In order to perform the operations efficiently, we split Hamiltonian into kinetic and potential parts as $\mathcal{H}=\mathcal{K}+\mathcal{V}$ and apply the Trotter-Suzuki decomposition:
\begin{equation}
 e^{-i\frac{\mathcal{H}\Delta t}{\hbar}}=e^{-i\frac{\mathcal{V}\Delta t}{2\hbar}}e^{-i\frac{\mathcal{K}\Delta t}{\hbar}}e^{-i\frac{\mathcal{V}\Delta t}{2\hbar}} + \mathcal{O}(\Delta t^3).
\end{equation} 
Since the effective mass of $\textrm{FaNDF}^0$ density functional is density independent, the kinetic part of the Hamiltonian $\mathcal{K}$ is diagonal in momentum representation and the operation $e^{-i\frac{\mathcal{K}\Delta t}{\hbar}}$ can be done efficiently by means of Fourier transforms.
The potential part has the following matrix structure,
\begin{equation}
 \mathcal{V}=\left(     \begin{array}{ll}
      U(\vect{r})  & \Delta(\vect{r}) \\
      \Delta^*(\vect{r}) & -U(\vect{r})
    \end{array} \right),
\end{equation} 
where the submatrices $U(\vect{r})$
(mean-field potential) and $\Delta(\vect{r})$
(pairing potential) are diagonal in the coordinate representation.  The exponent of
the potential part can be computed analytically,
\begin{multline}
 e^{-i\frac{\mathcal{V}\Delta t}{2\hbar}} =\\
 \left( 
\begin{array}{cc}
      \cos(\frac{\epsilon\Delta t}{2\hbar})-i\frac{U(\vect{r})}{\epsilon}\sin(\frac{\epsilon\Delta t}{2\hbar})  &  -i\frac{\Delta(\vect{r})}{\epsilon}\sin(\frac{\epsilon\Delta t}{2\hbar})\\
      -i\frac{\Delta^{*}(\vect{r})}{\epsilon}\sin(\frac{\epsilon\Delta t}{2\hbar}) & \cos(\frac{\epsilon\Delta t}{2\hbar})+i\frac{U(\vect{r})}{\epsilon}\sin(\frac{\epsilon\Delta t}{2\hbar})
    \end{array} 
 \right),
\end{multline} 
where
\begin{equation}
 \epsilon=\sqrt{U(\vect{r})^2 + |\Delta(\vect{r})|^2}.
\end{equation} 
The final complexity of the algorithm is governed by the FFT complexity. Note that this very efficient method cannot be used if spin-orbit term is included ($U(\vect{r})$ is no longer diagonal) or 
the effective mass is density dependent (kinetic part is no longer diagonal in momentum representation).

In the present simulations we have used an integration time step $\Delta t=0.054\fm/c$. This time step makes the time evolution stable within time intervals of about $18,000\fm/c$. 
Number of evolved wave-functions is about $37,000$ for neutrons and $16,000$ for protons. 

The integration is performed with an external time-dependent potential that couples only to protons,
\begin{equation}
 U_{\textrm{ext}}(\vect{r},t) = -\frac{1}{Z}\vect{F}_{\textrm{ext}}(t)\cdot\vect{r},
\end{equation}
where $Z$ is the number of protons. According to the Ehrenfest's theorem,
this external potential corresponds to an external force,
\begin{equation}
 \dfrac{d\left\langle \hat{\vect{p}}\right\rangle }{dt}=-\left\langle\vect{\nabla} U_{\textrm{ext}}(\vect{r},t)\right\rangle=\vect{F}_{\textrm{ext}}(t),
\end{equation} 
which is constant in space. The force is dynamically adjusted during the time evolution in such a way that the resulting motion of protons occurs at a constant velocity.

\begin{figure}[b]
\includegraphics[width=1.0\columnwidth, trim=30 60 20 35, clip]{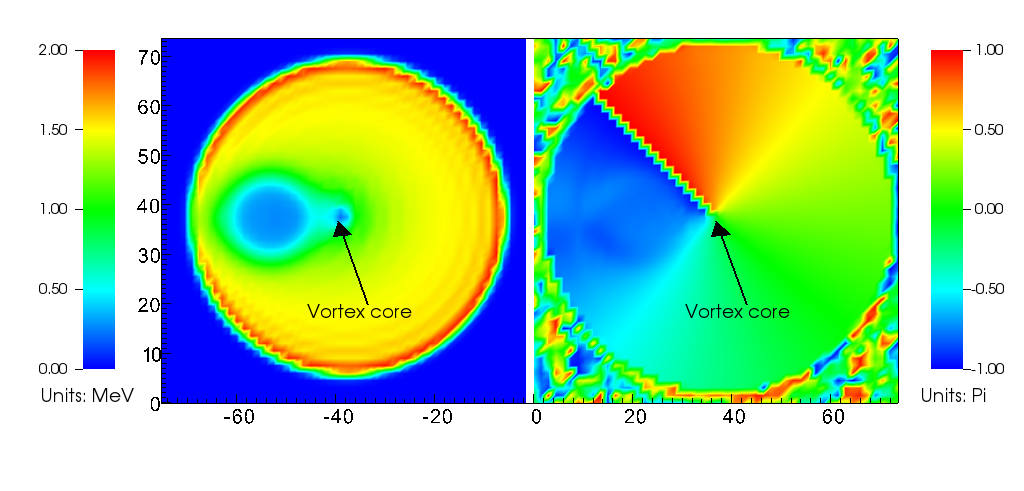}\\
\includegraphics[width=1.0\columnwidth, trim=30 60 20 35, clip]{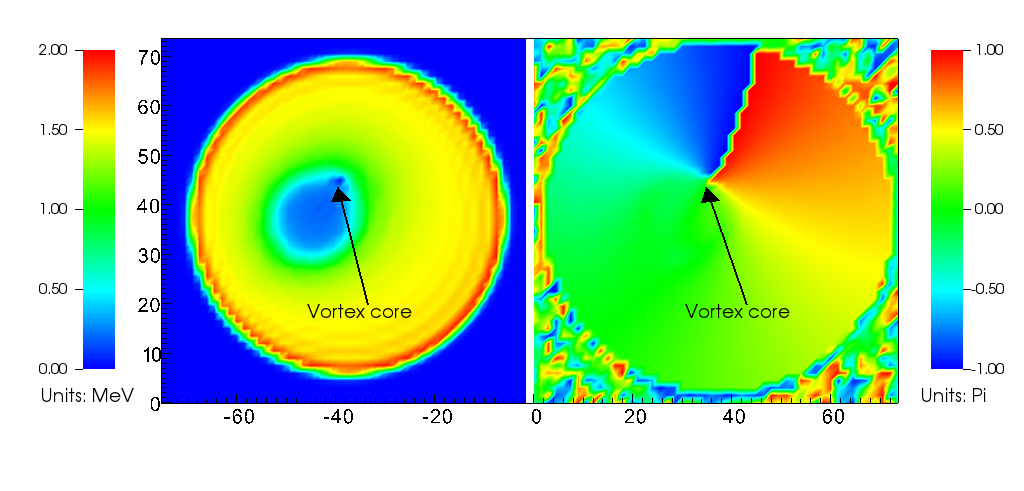}
\caption{ (Color online) Sections along XY-plane at $z=30\fm$ showing absolute values (left column) and phase (right column) of the pairing potential for two different vortex-nucleus configurations for background neutron density, $n=0.031\fm^{-3}$. Strong suppression of absolute value of the pairing potential is due to presence of the quantum vortex core and the nucleus. The phase of the pairing potential is almost unaffected by the presence of the nucleus, and it can be a clear signal of the vortex core position. Arrow shows detected vortex core position from analysis of the phase.
\label{fig:vdetect}}
\end{figure}

\section{Vortex detection}

In order to extract the vortex core location, we have analyzed the pairing field $\Delta$. Both the absolute value and the phase of the pairing field has been taken into account. It is known that in the vortex core the pairing field vanishes, while its phase rotates around this singular point. In our case, it is difficult to obtain precise information on the vortex core position from the location where $|\Delta|$ vanishes, especially in the case of the vortex being close to the nucleus. Because of high neutron density inside the nucleus, the absolute value of pairing gap is very small in this region. Therefore it is difficult to numerically locate the vortex core inside the nucleus only by analyzing the behavior of the pairing gap magnitude. The procedure has to be supplemented
by analysis of the pairing phase which allows to identify the vortex core position unambiguously. We found that the phase is weakly affected by the presence of the nucleus, as shown in Fig.~\ref{fig:vdetect}.

Thus, for each XY-plane we searched for a point around which the phase rotates by $2\pi$ and we attributed it to the vortex core position. Next, we created a line connecting vortex cores in each XY-plane using spline interpolation. 
The vortex core position for XY-plane crossing the center of mass of protons has been used to define the vortex-nucleus distance $R$. 

\section{Vortex tension}
\begin{figure}[t]
\includegraphics[width=1\columnwidth]{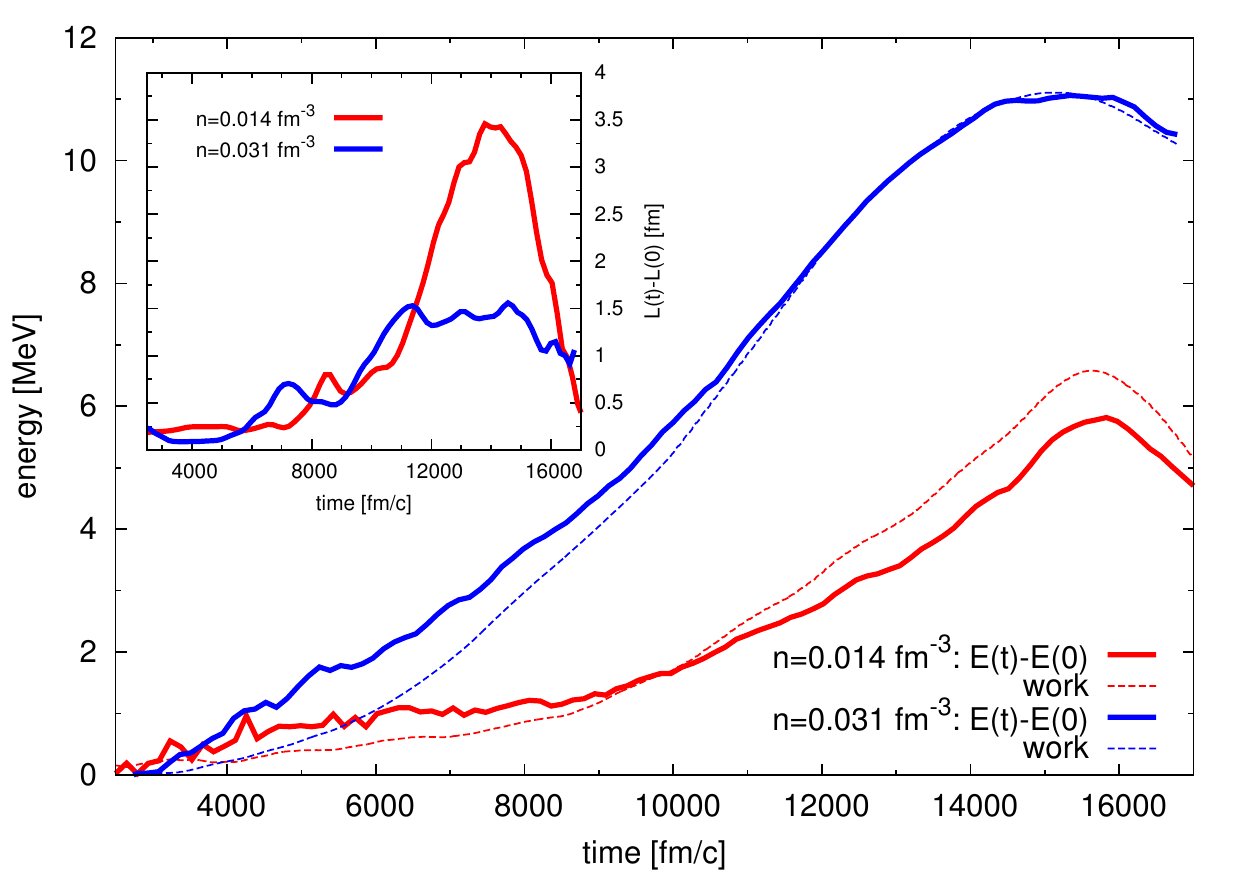}
\caption{ (Color online) Excitation energy $E(t)-E(0)$ of the system as a function of time. By dashed line work performed by external force computed by formula $W(t)=\int_{0}^{t}\vect{F}_{\textrm{ext}}(t^\prime)\cdot\vect{v}(t^\prime)\;dt^\prime$ is presented. In inset change of vortex length $L(t)-L(0)$ is shown as function of time.
\label{fig:vtension}}
\end{figure}

The time-dependent simulations allow for estimation of the vortex tension $T$ originating from the quantized superflow around the core. 
We have evaluated the excitation energy of the system $E^*(t)=E(t)-E(0)$ as a function of time due to the motion of the impurity. It is related to work performed by the external force $\vect{F}_{\textrm{ext}}(t)$. 
In Fig.~\ref{fig:vtension} we present $E^*(t)$ and compare it with the work computed according to the formula $W(t)=\int_{0}^{t}\vect{F}_{\textrm{ext}}(t^\prime)\cdot\vect{v}(t^\prime)\;dt^\prime$, where $\vect{v}(t)$ is the velocity of the center of mass of the protons. Indeed these two quantities agree with reasonable accuracy. The energy reaches maximum at the configuration corresponding to the vortex having the largest length, see inset of Fig.~\ref{fig:vtension}. If we assume that the total excitation energy is absorbed by the vortex, we can estimate the vortex tension as $T\simeq E^*(t_{\textrm{max}})/\Delta L_{\textrm{max}}$, where $t_{\textrm{max}}$ 
corresponds to the time when the vortex achieves the maximum length $L_{\textrm{max}}=L(0)+\Delta L_{\textrm{max}}$. Consequently we have extracted 
values $T\approx 5/3.5=1.4\;\textrm{MeV/fm}$ and $T\approx 11/1.5=7.3\;\textrm{MeV/fm}$ for densities $n=0.014\fm^{-3}$ and $0.031\fm^{-3}$, respectively. 
The results admit that the vortex is much stiffer for higher density case. Note that our estimation should be treated as an upper limit for 
the tension as some fraction of the excitation energy is absorbed by other degrees of freedom related for example to the nucleus, giving rise to the shape deformation. 
On the other hand such a quantity is more useful than the bare tension, as it takes into account
other degrees of freedom whose change may not be easy to disentangle in simplified models describing the vortices in the presence of nuclear impurities.

\section{Vortex-impurity interaction from irrotational and incompressible hydrodynamics}

\begin{figure}[t]
\includegraphics[width=1.15\columnwidth]{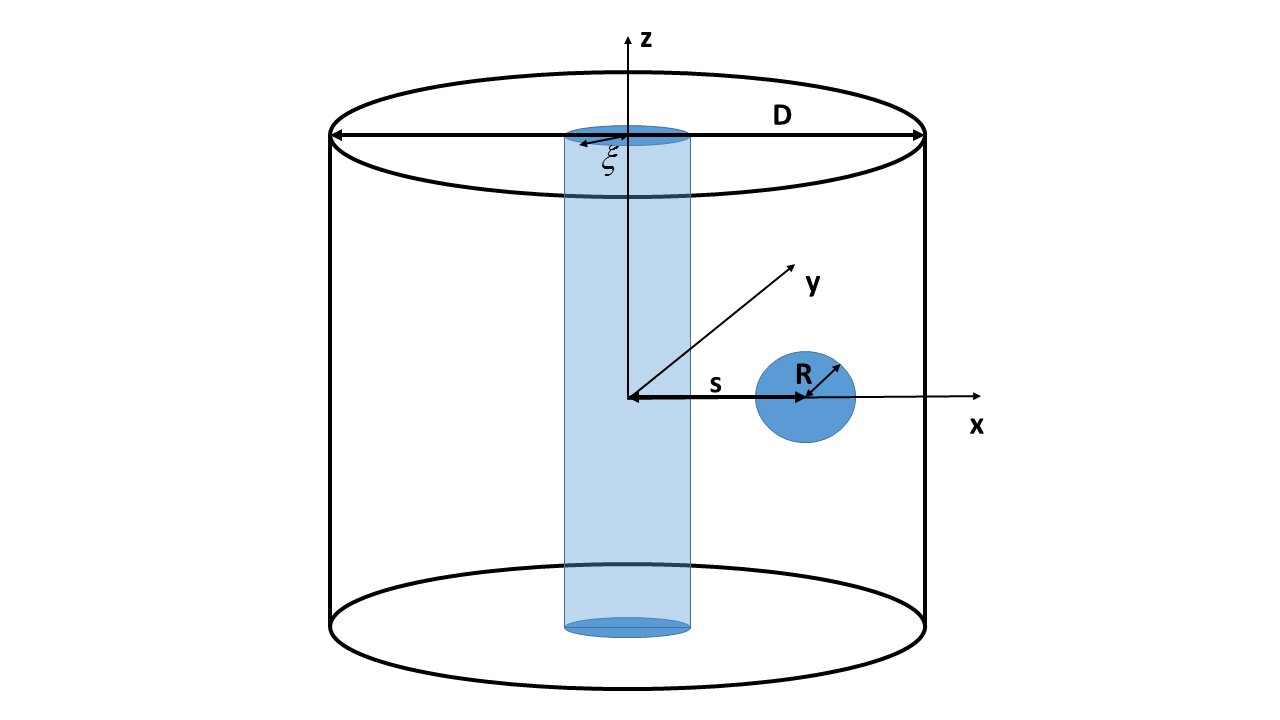}
\caption{ (Color online) Schematic picture showing the mutual arrangement of the vortex (of radius $\xi$) and the impurity (of radius $R$) together with
a cylinder of diameter $D$ ($D\rightarrow\infty$) defining the boundary of the system. 
\label{vortex_impurity}}
\end{figure}

The vortex impurity interaction can be extracted from irrotational and incompressible hydrodynamics following Refs. \cite{LambSM,EpsteinBaymSM, MagierskiBulgacSM,MagierskiSM}.
Within this simplified picture the interaction originates from the distortion of the motion of the superfluid component induced by the presence of the impurity.
Using the method of images one can construct a series of corrections to the velocity field. The convergence of the series
depends on the ratio between the radius of the impurity and the vortex-impurity distance, which play the role of the expansion parameter.
If the ratio is sufficiently small one may extract the leading effect, taking into account the first term of the series only. Namely, 
let us assume that the vortex is located at $x=y=0$ along the $z$ axis. Consequently its velocity potential reads: $\frac{\kappa}{2\pi} \phi$, where $\kappa =\oint\vect{v}\cdot d\vect{r}=2\pi\hbar/2m_n $ corresponds to the circulation,
while $\phi$ is the angle in XY-plane. The impurity of radius $R$ is located at the distance $s$ along $x$ axis (see Fig.~\ref{vortex_impurity}). The impurity is defined
as having a superfluid density inside denoted by $\rho_{in}$. In addition the impurity may have an instantaneous arbitrary velocity $\vect{u}$.
The presence of the impurity modifies the velocity potential:
\begin{equation}
 \Phi_{out}(r, \phi)=\frac{\kappa}{2\pi}\phi+A\frac{(\vect{r}-\vect{s})\cdot (\frac{\kappa}{2\pi s}\vect{e}_{y}-\vect{u})}{|\vect{r}-\vect{s}|^{3}},
\end{equation} 
where $\vect{e}_{y}$ is the unit vector along the $y$ axis, and $A$ is a constant.
The term proportional to $A$ represents the leading correction associated with the presence of the impurity, as it is
of the same order $1/s$ at the impurity boundary as the first term. The next order term which would represent
the correction due to the presence of the vortex is of the order $1/s^2$ and is neglected here.
This velocity potential should match the velocity potential inside the impurity:
\begin{equation}
 \Phi_{in}(r, \phi)=B_{0} + B_{1}(\vect{r}-\vect{s})\cdot\frac{\kappa}{2\pi s}\vect{e}_{y}+B_{2}(\vect{r}-\vect{s})\cdot\vect{u} ,
\end{equation} 
where $B_{i}$ are constants which can be determined from the boundary conditions at the surface of the impurity:
\begin{eqnarray}
\Phi_{in}|_{R}&=&\Phi_{out}|_R , \\
\rho_{in} ( \frac{\partial\Phi_{in}}{\partial r}-\vect{u} )|_R &=& \rho_{out} ( \frac{\partial\Phi_{out}}{\partial r}-\vect{u} )|_R ,
\end{eqnarray}
where the second equation is simply the continuity equation at the surface of the impurity.
The above conditions lead to the following expression for the velocity field $\Phi_{in}$:
\begin{equation}
 \Phi_{in}(r, \phi)=(1+\frac{A}{R^{3}}) (\vect{r}-\vect{s})\cdot\frac{\kappa}{2\pi s}\vect{e}_{y}+\frac{A}{R^{3}}(\vect{r}-\vect{s})\cdot\vect{u} ,
\end{equation} 
 Consequently the total energy of the system reads:
\begin{equation}
E = \frac{1}{2}\rho_{in}\int_{V_{i}} (\nabla\Phi_{in})^{2} d^{3}r + \frac{1}{2}\rho_{out}\int_{V-V_{i}-V_{vor}} (\nabla\Phi_{out})^{2} d^{3}r ,
\end{equation}
where $V_{i}$ and $V_{vor}$ represent the volumes of the impurity and the vortex, respectively. The volume $V$ denotes the volume
of the cylinder shown in the Fig.~\ref{vortex_impurity}.

The straightforward although lengthy calculations give the following expression for the total energy:
\begin{eqnarray}\label{hydro}
E &=& \frac{1}{4\pi}\rho_{out}\kappa^{2} H \ln \left (\frac{D}{2\xi} \right )
  + \frac{1}{2}\left ( \frac{4\pi}{3} R^{3} \frac{ (\rho_{out}-\rho_{in})^{2} }{ 2\rho_{out}+\rho_{in} } \right ) u^{2} \nonumber \\
  &+& \left ( 2\pi R^{3}\frac{\rho_{out}(\rho_{in}-\rho_{out})}{2\rho_{out}+\rho_{in}} \right ) \left ( \frac{\kappa}{2 \pi s} \right )^{2} ,
\end{eqnarray}
where $H$
is the height of the cylinder ($V=\pi D^{2}H/4$).
The first term represents the logarithmically divergent contribution of the
vortex to the energy, the second term corresponds to the kinetic energy of the
impurity, whereas the third term is the leading order contribution to the
vortex-impurity interaction (neglecting terms $1/s^3$
and higher).  Note that the effective mass agrees with the result in
Ref.~\cite{MagierskiSM}.  The obtained result indicates that the force between
the vortex and impurity is attractive if $\rho_{in} < \rho_{out}$
and repulsive if $\rho_{in} > \rho_{out}$,
and its leading component behaves like $1/s^{3}$
as a function of the distance.  This result requires a comment. In the case
when $\rho_{in}=0$
it has been derived in Ref.~\cite{EpsteinBaymSM} providing an argument for the
existence of the pinning force. However more general treatment indicates that
the anti-pinning effect can appear as well.  It is difficult to relate
superfluid densities $\rho_{in}$
and $\rho_{out}$
to the densities originating from the microscopic description provided by DFT.
Therefore this result should be treated as providing only an asymptotic
expression (when $R/s\rightarrow 0$)
for the real vortex-impurity interaction, and the sign as well as the magnitude
of the coefficient, responsible for the intensity of the force, has to be
determined from the microscopic approach. Our parametrization of the pairing
interaction underestimates a little the strength of the pairing gaps inside
nuclei (where both protons and neutrons are present) as inferred from
phenomenological studies. However, the magnitude of the pairing gap is still a
matter of debate~\cite{CaoSM,GandolfiSM}, and especially for the case of a nucleus
embedded in neutron matter it is completely unknown.  Nuclei embedded in a
neutron superfluid swell, and thus their pairing properties would tend to be
more similar to neutron matter~\cite{UmarSM}.  Since we possibly underestimate
the pairing gaps inside nuclei, the neutron superfluid fraction is only
underestimated in the worst case, if one were to use the hydrodynamic
calculation~\eqref{hydro}.  Therefore the repulsive force between the vortex
and the nucleus is underestimated as well. The previous hydrodynamic
calculation of the vortex-nucleus interaction~\cite{EpsteinBaymSM} assumed that
$\rho_{in}\equiv 0$,
which resulted in an attractive force, not repulsive as we obtain.  The
character of the pairing in a nucleus immersed in neutron superfluid is not
known from ab initio calculations

The first term in Eq.~(\ref{hydro}) represents the energy of the vortex $E_{vor}$ inside the cylinder and can be used to calculate 
the vortex tension $T=\frac{d E_{vor}}{d H}$ characterizing the stiffness of the vortex. The formula allows to determine the vortex tension as a function of superfluid (neutron matter) density $\rho_{s}$. Namely: $T=\frac{1}{4\pi}\rho_{s}\kappa^{2}\ln\frac{D}{2\xi}$, where $D$ is the diameter
of the Wigner-Seitz cell and the size of the vortex core $\xi$ is related to the coherence length \cite{SevesoSM}. 

\section{Force per unit length - fitting details}

\begin{figure}[t]
\includegraphics[width=1\columnwidth]{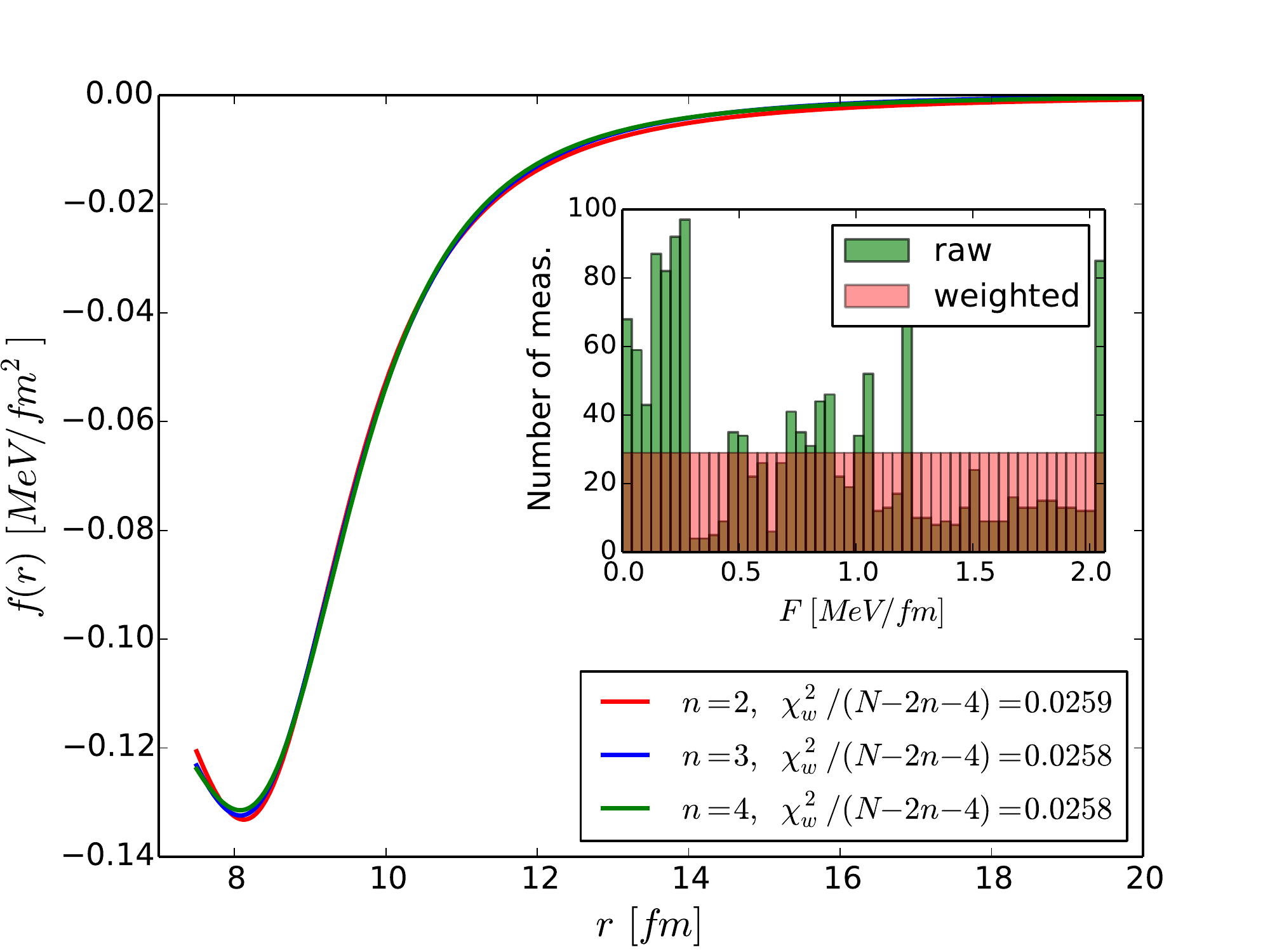}
\caption{ (Color online) Force per unit length for different orders of Pad\'{e} approximant for density $0.014\fm^{-3}$. Weighted $\chi^2$ per degree of freedom saturates at order $n=2$. In inset (raw) histogram of measured forces $|\vect{F}_i|$  is shown. After introducing weights the histogram becomes uniform.
\label{fig:force_f_comparison}}
\end{figure}

In order to extract the force per unit length $f(r;\{a_k,b_k\})$ we minimized weighted $\chi^2$ quantity of the form,
\begin{equation}
 \chi^2_w=\sum_{i=1}^{N} w(|\vect{F}_i|)\left( \vect{F}_i - \vect{F}_i^{(f)}(\{a_k,b_k\})\right)^2,
\end{equation} 
with respect to parameters $\{a_k,b_k\}$, where $i$ counts set of measurements (frames from movies), $\vect{F}_i$ is measured total force for each measurement, and $\vect{F}_i^{(f)}$ is the predicted total force,
\begin{equation}
 \vect{F}_i^{(f)}=\int_{L_i}f(r;\{a_k,b_k\})\sin\alpha\;\vect{e}_{r}\,dl.
 \label{eqn:F_theory}
\end{equation} 
(See the main text for meaning of symbols in the above equation). 
To parametrize $f(r)$ we have used Pad\'{e} approximant
\begin{equation}
 f(r) = \dfrac{\sum_{k=0}^{n}a_k r^k}{1+\sum_{k=1}^{n+3}b_k r^k},\label{eqn:f_r_Pade}
\end{equation} 
where we imposed asymptotic behavior $f(r\rightarrow\infty)\propto 1/r^{3}$,
predicted by irrotational, incompressible hydrodynamics. Number of fitting parameters is $2n+4$.
Our set of measurements contain mostly configurations with significant separations $R$, where the vortex-nucleus interaction is weak. To prevent these measurements to dominate $\chi^2$ we introduced weights $w(|\vect{F}_i|)$, in such a way to get the uniformly weighted histogram of measured forces $|\vect{F}_i|$, see inset of Fig.~\ref{fig:force_f_comparison}. We have checked the influence of the weighting form $w$ on the final result and we found that this dependence is very weak.

\begin{figure}[t]
\includegraphics[width=\columnwidth]{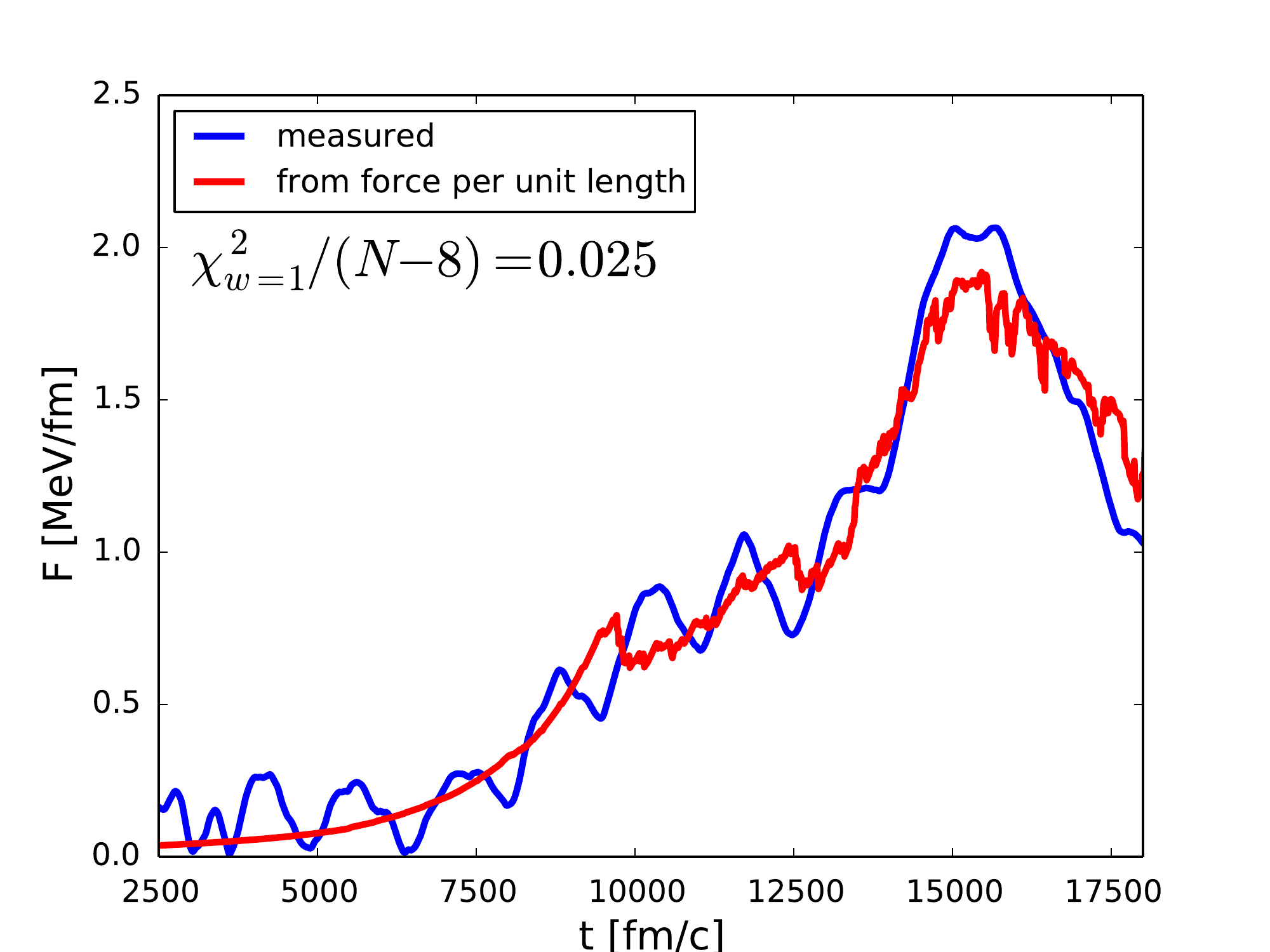}\\
\includegraphics[width=\columnwidth]{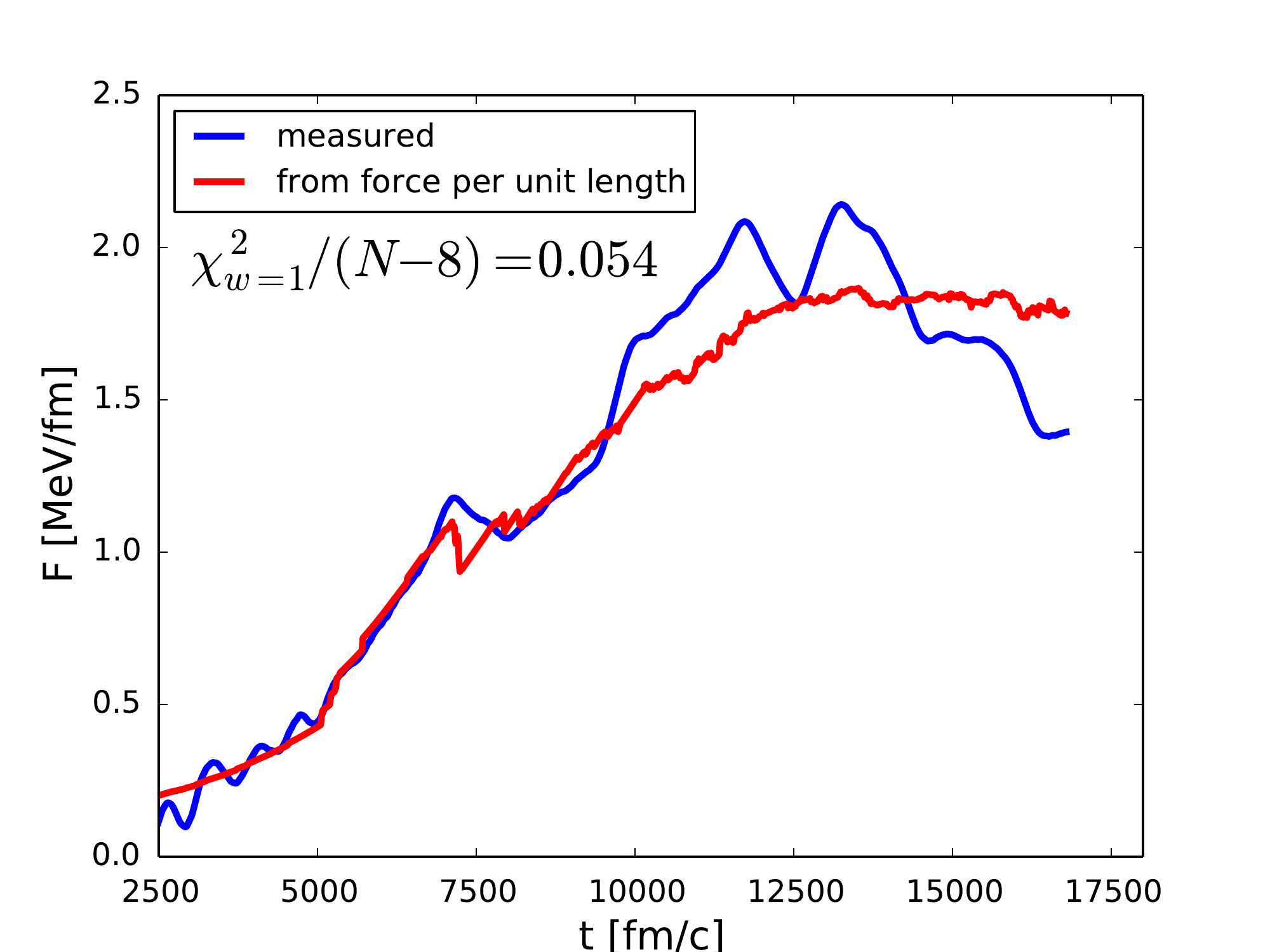}
\caption{ (Color online) Quality of the total force $|\vect{F}|$ reproduction as a function of time for $n=0.014\fm^{-3}$ (top) and $n=0.031\fm^{-3}$ (bottom).
Value of unweighted chi-square per degree of freedom is also given.
\label{fig:force_reproduction}}
\end{figure}

We have performed minimization for different orders of Pad\'{e} approximants and we found that negligible improvements of $\chi^2_w$ are gained for orders higher than $n=2$. Weighted chi-square per degree of freedom saturates at values $\chi^2_w/(N-2n-4)\cong 0.026$ and $0.038$ for densities $0.014\fm^{-3}$ and $0.031\fm^{-3}$ respectively. The corresponding unweighted (all $w(|\vect{F}_i|)=1$) values  are $\chi^2_{w=1}/(N-2n-4)\cong 0.025$ and $0.054$.
In Fig.~\ref{fig:force_f_comparison} we demonstrate the convergence of Pad\'{e} approximant for different orders ($n=2,3,4$) for lower density case.
From the figure, good convergence of the Pad\'{e} approximant can be seen.
The fit is meaningful only for $r\gtrsim8\fm$ and $r\gtrsim9\fm$ for densities $n=0.014\fm^{-3}$ and $n=0.031\fm^{-3}$, respectively, and thus, 
the appearing minimum of the force per unit length for small separations has no meaning. In Table~\ref{tab:PadeCoeff} we provide values of coefficients $a_k$ and $b_k$ for $n=2$. 
\begin{table}[b]
  \begin{tabular}{ | l || r | r |}
    \hline
     & $n=0.014\fm^{-3}$ & $n=0.031\fm^{-3}$\\
     \hline\hline
$a_0\;[\MeV\fm^{-2}]$ & -4549.83 & -3735.28 \\
$a_1\;[\MeV\fm^{-3}]$ & -4525.79 & -988.42 \\
$a_2\;[\MeV\fm^{-4}]$ & -505.60 & -4257.99 \\
\hline
$b_1\;[\fm^{-1}]$ & 6455.46 & 6738.74 \\
$b_2\;[\fm^{-2}]$ & 6299.41 & 8430.92 \\
$b_3\;[\fm^{-3}]$ & 23440.34 & 35498.57 \\
$b_4\;[\fm^{-4}]$ & -5640.24 & -7190.64 \\
$b_5\;[\fm^{-5}]$ & 341.73 & 397.51 \\
     \hline
  \end{tabular}
  \caption{Coefficients of Pad\'{e} approximant for $n=2$.}
  \label{tab:PadeCoeff}
\end{table}
To show the quality of reproduction of the total force, we have shown in Fig.~\ref{fig:force_reproduction} the measured force $\vect{F}_i$ (blue solid line) and the reconstructed force $\vect{F}_i^{(f)}$, according to Eq.~(\ref{eqn:F_theory}) with Pad\'{e} approximants $n=2$ as a function of time. Top (bottom) panel shows results for background neutron density, $n=0.014\fm^{-3}$ ($0.031\fm^{-3}$).

In order to quantify goodness of the fit one should consider reduced chi-squared statistic
\begin{equation}
\chi^2_{\mathrm{red.}}=\dfrac{1}{N-2n-4}\sum_{i=1}^{N}\dfrac{\left( \vect{F}_i - \vect{F}_i^{(f)}\right)^2}{\sigma_i^2},
\end{equation}
where $\sigma_i$ is uncertainty of $i$-th measurement. Fit is considered as meaningful if $\chi^2_{\mathrm{red.}}\approx 1$. Oscillations of the measured force seen in Fig.~\ref{fig:force_reproduction} clearly demonstrate that the measurements are affected by uncertainties. (Their source is discussed in next section.) They are of size $0.2-0.3\;\mathrm{MeV/fm}$ and it can be used as estimate of the measurement uncertainty $\sigma$. Then we can estimate the reduced statistics 
\begin{equation}
\chi^2_{\mathrm{red.}}\cong\dfrac{1}{N-2n-4}\dfrac{\chi^2_{w=1}}{\sigma^2}.
\end{equation}
For both densities we obtain value close to one. It confirms that the fit is meaningful at statistical level.

\section{Velocity of the nucleus}

\begin{figure}[t]
\includegraphics[width=0.85\columnwidth]{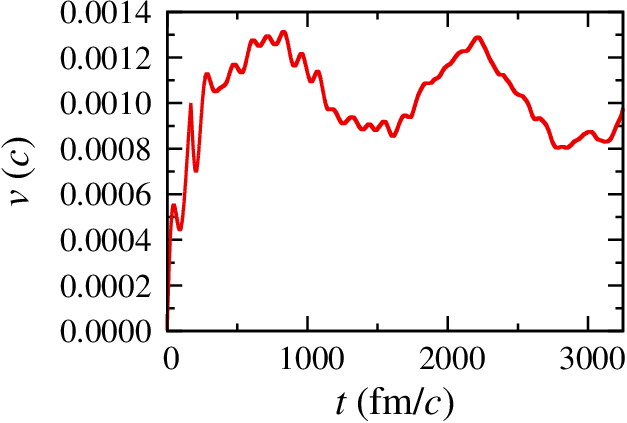}
\caption{(Color online)
Velocity of center of mass of protons as a function of time for
a state with $n=0.014\fm^{-3}$ without vortex. See text for details.
}
\label{fig:velocity_wo_vortex}
\end{figure}

The measured force $\vect{F}_i$ shown in Fig.~\ref{fig:force_reproduction} exhibits
noticeable fluctuations with a period as large as 2,000~fm/$c$. In order to check
the origin of the fluctuations, we have performed the following simulation:
starting from the state where nucleus ($Z=50$) is immersed in a neutron
superfluid ($n=0.014\fm^{-3}$) without vortex, we accelerated the nucleus
along $z$ axis  up to $v=0.001\,c$.
Then, we switched off the external potential. In Fig.~\ref{fig:velocity_wo_vortex}, the velocity
of center of mass of protons is shown as a function of time. In the ideal
situation where no energy transfer to internal degrees of freedom is present, the velocity should
stay constant. Indeed, as shown in Fig.~\ref{fig:velocity_wo_vortex},
we observed the nuclear motion with a roughly constant velocity.
However, we found that the velocity is also fluctuating about the value $v=0.001\,c$
with a similar period as that observed in the measured force shown
in Fig.~\ref{fig:force_reproduction}. We concluded therefore that the fluctuations
are due to excitations of a background neutron superfluid induced by
the motion of the nucleus.
We expect that the fluctuations will disappear in the limit of $v \rightarrow 0$. We would like to emphasize that the conclusions of this paper are not affected by the fluctuations.

We note that the dragging velocity $v=0.001\,c$ is
far below the critical velocity of the system and is sufficiently small that the systems follow an almost adiabatic path:
for example, after dragging the nucleus for a
time of $17,000\fm/c$,
we increased the energy per particle by less than $0.2\;\textrm{keV}$.

\section{Movies}
We provide movies showing dynamics of the system. Elements shown in the movies are:
\begin{description}
 \item[Red dot] Position of the center of mass of protons.
 \item[Blue line] Position of the quantum vortex core.
 \item[Black vector attached to Red dot] Measured vortex-nucleus force $\vect{F}$.
 \item[Green vector attached to Red dot] Reconstructed vortex-nucleus force from extracted force per unit length.
 \item[Green vectors attached to Blue line] Contributions to the reconstructed force generated by each vortex element. They are plotted with minus sign and multiplied by factor 3 for better visibility.
 \item[Dashed red line] Shape of the nucleus defined as points where density of protons drops to value $0.005\fm^{-3}$.
 \item[Blue triangles on XY-plane] Trajectory of central element of the vortex up to given time is shown. 
 \item[Label $F_m(R)$ ] Absolute value of the measured force for a given frame. In parenthesis distance $R$ between the center of mass of protons and the quantum vortex (as defined in section \textit{Vortex detection}) are shown. 
 \item[Label $F_r(R)$] Absolute value of the force computed from the extracted force per unit length.
 \item[Label $Q$] Quadrupole moment of proton density distribution. 
\end{description}
For better visibility projections of main view along frame axes are depicted on sides of the box.\\
\\
List of movies:
\begin{enumerate}
 \item File: \verb|n0_014_unpinned.mp4|\\
 Dynamics of the system with background neutron density $n=0.014\fm^{-3}$ and a nucleus consisting of $50$ protons. Initial state: unpinned configuration (top left panel of Fig.~\ref{fig:initstates_frames}). Algorithm for the external force adjustment needs about $2,500\fm/c$ to put the nucleus into constant velocity movement. Only after this time, the force can be attributed as vortex-nucleus force.\\
 YouTube: \url{https://youtu.be/yIS3I36wQ9U}
 \item File: \verb|n0_031_unpinned.mp4|\\
 Dynamics of the system with background neutron density $n=0.031\fm^{-3}$ and a nucleus consisting of $50$ protons. Initial state: unpinned configuration (bottom left panel of Fig.~\ref{fig:initstates_frames}).\\
 YouTube: \url{https://youtu.be/rFP7Tbh1mVQ}
  \item File: \verb|n0_014_pinned.mp4|\\
 Dynamics of the system with background neutron density $n=0.014\fm^{-3}$ and a nucleus consisting of $50$ protons. Initial state: pinned configuration (top right panel of Fig.~\ref{fig:initstates_frames}). The force is not shown as our algorithm for the external force adjustment needs about $2,500\fm/\textrm{c}$ to put the nucleus into constant velocity movement, and obtained force is not vortex-nucleus force.\\
 YouTube: \url{https://youtu.be/qdQHj4wjkv8}
 \item File: \verb|n0_031_pinned.mp4|\\
 Dynamics of the system with background neutron density $n=0.031\fm^{-3}$ and a nucleus consisting of $50$ protons. Initial state: pinned configuration (bottom right panel of Fig.~\ref{fig:initstates_frames}).\\
 YouTube: \url{https://youtu.be/DRRNJey-lBg}
 \item File: \verb|n0_014_unpinned_with_f.mp4|\\
 Movie demonstrating quality of the total force reconstruction by the force per unit length for simulation \verb|n0_014_unpinned.mp4|.\\
 YouTube: \url{https://youtu.be/Eb459PyyS4A}
 \item File: \verb|n0_031_unpinned_with_f.mp4|\\
 Movie demonstrating quality of the total force reconstruction by the force per unit length for simulation \verb|n0_031_unpinned.mp4|.\\
 YouTube: \url{https://youtu.be/avlk4vZuGrM}
\end{enumerate}~

In addition, we provide a movie showing response of spinning gyroscope 
when pushed. It demonstrates that rotating object moves perpendicular to
the external force and direction of the response is set by $\vect{\Omega}\times \vect{F}$, where
$\vect{\Omega}$ is angular velocity and $\vect{F}$ is external force.\\
File: \verb|gyroscope.mp4|\\
YouTube: \url{https://youtu.be/iYLjmC7fpzk}


\begin{thebibliography}{99}

\bibitem{Radhakrishnan:1969}%
  V.~Radhakrishnan and R.~N.~Manchester, %
  Detection of a change of state in the pulsar {PSR} 0833-45, %
  \href{\doibase 10.1038/222228a0}{Nature \textbf{222}, 228--229 (1969).}
\bibitem{Reichley:1969}%
  P.~E.~Reichley and G.~S.~Downs, %
  Observed decrease in the periods of pulsar PSR 0833-45, %
  \href{\doibase 10.1038/222229a0}{Nature \textbf{222}, 229--230(1969)}.
\bibitem{LattimerPrakash} J.M. Lattimer, M. Prakash, The Physics of Neutron
  Stars, Science {\bf 304}, 536-542 (2004).
\bibitem{AndersonItoh} P.W~Anderson, N~Itoh, Pulsar glitches and restlessness
  as a hard superfluidity phenomenon, Nature {\bf 256}, 25 (1975)
\bibitem{ModelsGlitch} B. Haskell, A. Melatos, Models of pulsar glitches,
  Int. J. Mod. Phys. D {\bf 24}, 1530008 (2015).
\bibitem{Pizzochero2} P. M. Pizzochero, L. Viverit, R. A. Broglia,
  Vortex-nucleus interaction and pinning forces in neutron stars,
  Phys. Rev. Lett. {\bf 79}, 3347 (1997).
\bibitem{Donati} P. Donati, P.M. Pizzochero, Is there Nuclear Pinning of
  Vortices in Superfluid Pulsars, Phys. Rev. Lett. {\bf 90}, 211101 (2003).
\bibitem{Donati2} P. Donati, P.M. Pizzochero, Realistic energies for vortex
  pinning in intermediate-density neutron star matter, Phys. Lett. B {\bf 640},
  74 (2006).
\bibitem{Donati3} P. Donati, P.M. Pizzochero, Fully consistent semi-classical
  treatment of vortex–nucleus interaction in rotating neutron stars,
  Nucl. Phys. A. {\bf 742}, 363 (2004).
\bibitem{Seveso} S. Seveso, P. M. Pizzochero, F. Grill and B. Haskell,
  Mesoscopic pinning forces in neutron star crusts, MNRAS {\bf 455}, 3952
  (2016).
\bibitem{Avogadro1} P. Avogadro, F. Barranco, R. A. Broglia, and E. Vigezzi,
  Quantum calculation of vortices in the inner crust of neutron stars,
  Phys. Rev. C 75, 012805(R) (2007).
\bibitem{Avogadro2} P. Avogadro, F. Barranco, R.A. Broglia, E. Vigezzi,
  Vortex-nucleus interaction in the inner crust of neutron stars, Nucl. Phys. A
  {\bf 811}, 378 (2008).
\bibitem{Pizzochero1} P. M. Pizzochero, Pinning and Binding Energies for
  Vortices in Neutron Stars: Comments on Recent Results, (2007),
  arXiv:0711.3393.
\bibitem{COCG} S. Jin, A. Bulgac, K. Roche, G. Wlaz\l{}owski, 
Coordinate-Space Solver for Superfluid Many-Fermion Systems with Shifted Conjugate Orthogonal Conjugate Gradient Method, (2016), arXiv:1608.03711 [nucl-th].
\bibitem{BulgacPinning} A. Bulgac, M.M. Forbes, R. Sharma, Strength of the
  Vortex-Pinning Interaction from Real-Time Dynamics, Phys. Rev. Lett. {\bf 110}, 241102 (2013).
\bibitem{Supplemental} See Supplemental Material at \{URL will be provided by
  the publisher \} which includes Refs.~\cite{QuantumFrictionAA,RTakayamaAA,SYamamotoAA,GandolfiAA,LambAA,MagierskiBulgacAA,MagierskiAA,CaoAA,UmarAA} 
  for discussion of technical aspects including generation of
  initial configurations, integration of TDSLDA equations, vortex detection
  algorithm, force per unit length fitting procedure and list of accompanying   movies.
\bibitem{BLink} B. Link, Dynamics of Quantum Vorticity in a Random Potential, Phys. Rev. Lett. {\bf 102}, 131101 (2009).
\bibitem{Quantvor} {\em Quantized Vortex Dynamics and Superfluid Turbulence},
  Lecture Notes in Physics, Eds. C.F. Barenghi, R.J. Donnelly, W.F. Vinen,
  Springer-Verlag Berlin Heidelberg 2001.
\bibitem{Epstein} R.I.~Epstein, G.~Baym , Vortex drag and the spin-up time
  scale for pulsar glitches, Astrophys. J. {\bf 387}, 276 (1992).
\bibitem{Baym} R.I. Epstein and G. Baym, Vortex pinning in neutron stars, Ap. J. {\bf 328}, 680 (1988).  
\bibitem{Antonelli} M. Antonelli, P. Pizzochero, Axially symmetric equations
  for differential pulsar rotation with superfluid entrainment,
  arXiv:1603.02838 (2016).
\bibitem{PRL__2009} A. Bulgac and S. Yoon, %
  Large Amplitude Dynamics of the Pairing Correlations in a Unitary Fermi
  Gas, %
  Phys. Rev. Lett. {\bf 102}, 085302 (2009).
\bibitem{Science__2011} A. Bulgac, Y.-L. Luo, P. Magierski, K. J. Roche, and
  Y. Yu, %
  Real-Time Dynamics of Quantized Vortices in a Unitary Fermi Superfluid, %
  Science, {\bf 332}, 1288 (2011).
\bibitem{LNP__2012} A. Bulgac, P. Magierski, and M.M. Forbes, %
  The Unitary Fermi Gas: From Monte Carlo to Density Functionals, in {\it
    BCS-BEC Crossover and the Unitary Fermi Gas}, %
  edited by W. Zwerger, Lecture Notes in Physics, Vol. 836, pp 305-373
  (Springer, Heidelberg, 2012).
\bibitem{PRL__2012} A. Bulgac, Y.-L. Luo, and K.J. Roche, %
  Quantum Shock Waves and Domain Walls in Real-Time Dynamics of a Superfluid
  Unitary Femi Gas, %
  Phys. Rev. Lett. {\bf 108}, 150401 (2012).
\bibitem{ARNPS__2013} A. Bulgac, Time-Dependent Density Functional Theory and
  Real-Time Dynamics of Fermi Superfluids, Ann. Rev. Nucl. Part. Sci. {\bf 63},
  97 (2013).
\bibitem{PRL__2014} A. Bulgac, M.M. Forbes, M.M. Kelley, K.J. Roche, and
  G. Wlaz\l{}owski, Quantized Superfluid Vortex Rings in the Unitary Fermi Gas,
  Phys. Rev. Lett. {\bf 112}, 025301 (2014).
\bibitem{PRA__2015} G. Wlaz\l{}owski, A. Bulgac, M. M. Forbes, and K. J. Roche,
  Life Cycle of Superfluid Vortices and quantum turbulence in the Unitary Fermi
  Gas, Phys. Rev. A {\bf 91}, 031602(R) (2015).
\bibitem{PRC__2011} I. Stetcu, A. Bulgac, P. Magierski, and K.J. Roche,
  Isovector Giant Dipole Resonance from 3D Time-Dependent Density Functional
  Theory for Superfluid Nuclei, Phys. Rev. C {\bf 84}, 051309(R) (2011).
\bibitem{PRL__2015} I. Stetcu, C.A.  Bertulani, A. Bulgac, P. Magierski, and
  K.J. Roche, %
  Relativistic Coulomb Excitation within Time-Dependent Superfluid Local
  Density Approximation, %
  Phys. Rev. Lett. {\bf 114}, 012701 (2015).
\bibitem{PRL__2016} A. Bulgac, P. Magierski, K.J. Roche, I. Stetcu, %
  Induced Fission of ${}^{240}$Plutonium
  within a Real-Time Microscopic Framework, %
  Phys. Rev. Lett. {\bf 116}, 122504 (2016).
\bibitem{Mag2016} P. Magierski, %
  Nuclear Reactions and Superfluid Time Dependent Density Functional Theory, to
  appear in {\em Progress of time-dependent nuclear reaction theory}, Betham
  Science Publishers 2016.
\bibitem{NegeleVautherin} J. Negele and D. Vautherin, Neutron star matter at
  sub-nuclear densities, Nucl. Phys. A {\bf 207}, 298 (1973).
\bibitem{Fayans1} S.A. Fayans, JETP Letters, %
  Towards a universal nuclear density functional, {\bf 68}, 169 (1998);
\bibitem{Fayans2} S.A. Fayans, S.V. Tolokonnikov, E.L. Trykov, D. Zawischa, %
  Nuclear isotope shifts within the local energy-density functional approach,
  Nucl. Phys. A {\bf 676}, 49 (2000)
\bibitem{FriedmanPandharipande} B. Friedman and V. R. Pandharipande, Hot and
  cold nuclear and neutron matter, Nucl. Phys. A {\bf 361}, 502 (1981)
\bibitem{Wiringa} R. B. Wiringa, V. Fiks, and A. Fabrocini, Equation of state
  for dense nucleon matter, Phys. Rev. C {\bf 38}, 1010 (1988).
\bibitem{PRL__2003a} Y. Yu and A. Bulgac, Energy Density Functional Approach to
  Superfluid Nuclei, Phys. Rev. Lett. {\bf 90}, 222501 (2003) and {\it Appendix
    to:} Energy Density Functional Approach to Superfluid Nuclei,
  nucl-th/0302007.
\bibitem{arxiv:1507} S.V. Tolokonnikov, I.N. Borzov, M. Kortelainen,
  Yu.S. Lutostansky, and E.E. Saperstein, Fayans functional for deformed
  nuclei. Uranium region, arXiv:1507.06607, EPJ Web of Conferences, {\bf 107},
  02003 (2016), and earlier references therein.
\bibitem{PRL__2003} A. Bulgac and Y.Yu, %
  Spatial Structure of a Vortex in Low Density Neutron Matter, %
  Phys. Rev. Lett. {\bf 90}, 161101 (2003).
\bibitem{PRL__2002} A. Bulgac and Y. Yu, %
  Renormalization of the Hartree-Fock-Bogoliubov Equations in the Case of a
  Zero Range Pairing Interaction,%
  Phys. Rev. Lett.{\bf 88}, 042504 (2002).
\bibitem{PRC__2002} A. Bulgac, %
  Local Density Approximation for Systems with Pairing Correlations, %
  Phys. Rev. C 65 051305(R) (2002).
\bibitem{PRL__2003b} A. Bulgac and Y. Yu, %
  The vortex state in a strongly coupled dilute atomic fermionic superfluid, %
  Phys. Rev. Lett. {\bf 91}, 190404 (2003).
\bibitem{PRA__2007} A. Bulgac, %
  Local Density Functional Theory for Superfluid Fermionic Systems: The Unitary
  Gas, %
  Phys. Rev. A {\bf 76}, 040502(R) (2007).
\bibitem{PRL__2008} A. Bulgac and M.M. Forbes, %
  A Unitary Fermi Supersolid: The Larkin-Ovchinnikov Phase, %
  Phys. Rev. Lett. {\bf 101}, 215301 (2008).
\bibitem{Sedrakian} A. Sedrakian, %
  Vortex repinning in neutron start crusts, %
  Mon. Not. R. Astron. Soc. {\bf 277}, 225 (1995).
\bibitem{MIT1} M.J.H. Ku, W. Ji, B. Mukherjee, E. Guardado-Sanchez,
  L. W. Cheuk, T. Yefsah, and M. W. Zwierlein, %
  Motion of a Solitonic Vortex in the BEC-BCS Crossover, %
  Phys. Rev. Lett. {\bf 113}, 065301 (2014).
\bibitem{MIT2} M.J.H. Ku, B. Mukherjee, T. Yefsah, and M.W. Zwierlein, %
  Cascade of Solitonic Excitations in a Superfluid Fermi gas: From Planar
  Solitons to Vortex Rings and Lines, %
  Phys. Rev. Lett. {\bf 116}, 045304 (2016).
  \bibitem{Andersson} N. Andersson, T. Sidery, and G.L. Comer, Superfluid neutron star turbulence, MNRAS {\bf 381}, 747 (2007).
\bibitem{VinenNiemela} W.F. Vinen, J.J. Niemela, %
  Quantum Turbulence, %
  J. Low. Temp. Phys. {\bf 128}, 167 (2002).
\bibitem{Tsubota} M. Tsubota, M. Kobayashi, H. Takeuchi, %
  Quantum hydrodynamics, Phys. Rep. {\bf 522}, 191 (2013).
\bibitem{QuantumFrictionAA} A. Bulgac, M. M. Forbes, K. J. Roche, and G. Wlaz\l{}owski, 
  Quantum Friction: Cooling Quantum Systems with Unitary Time Evolution, (2013), arXiv:1305.6891 [nucl-th].
\bibitem{RTakayamaAA} R. Takayama, T. Hoshi, T. Sogabe, S.-L. Zhang, T. Fujiwara, 
  Linear Algebraic Calculation of Green's function for Large-Scale Electronic Structure Theory, Phys. Rev. B {\bf 73}, 165108 (2006).
\bibitem{SYamamotoAA} S. Yamamoto, T. Sogabe, T. Hoshi, S.-L. Zhang, T. Fujiwara, 
  Shifted COCG method and its application to double orbital extended Hubbard model, J. Phys. Soc. Jpn. {\bf 77}, 114713 (2008).
\bibitem{GandolfiAA} S. Gandolfi, A. Gezerlis, and J.   Carlson, Neutron matter from low to high density, Annu. Rev. Nucl. Part. Sci. {\bf 65}, 303 (2015).
\bibitem{LambAA} H. Lamb, Hydrodynamics, Cambridge University Press (1975).
\bibitem{MagierskiBulgacAA} P. Magierski, A. Bulgac, Nuclear hydrodynamics in the inner crust of neutron stars, Acta Phys. Polon. B35, 1203 (2004).	
\bibitem{MagierskiAA} P. Magierski, In-medium ion mass renormalization and lattice vibrations in the neutron star crust, Int. J. Mod. Phys. E13, 371 (2004).
\bibitem{CaoAA} L.G. Cao, U. Lombardo, amd P. Schuck, Screening effects in superfluid nuclear and neutron matter within Brueckner theory, Phys. Rev. C {\bf 74}, 064301 (2006).
\bibitem{UmarAA} A.S. Umar, V.E. Oberacker, C.J. Horowitz, P.-G. Reinhard, J.A. Maruhn, Swelling of nuclei embedded in neutron gas and consequnces for fusion, Phys. Rev. C {\bf 92}, 025808 (2015).
\end{thebibliography}

\begin{thebibliography}{99}
\bibitem{EdisonSM} www.nersc.gov/systems/edison-cray-xc30/
\bibitem{QuantumFrictionSM} A. Bulgac, M. M. Forbes, K. J. Roche, and G. Wlaz\l{}owski, Quantum Friction: Cooling Quantum Systems with Unitary Time Evolution, (2013), arXiv:1305.6891 [nucl-th]
\bibitem{RTakayamaSM} R. Takayama, T. Hoshi, T. Sogabe, S.-L. Zhang, T. Fujiwara, Linear Algebraic Calculation of Green's function for Large-Scale Electronic Structure Theory, Phys. Rev. B {\bf 73}, 165108 (2006).
\bibitem{SYamamotoSM} S. Yamamoto, T. Sogabe, T. Hoshi, S.-L. Zhang, T. Fujiwara, Shifted COCG method and its application to double orbital extended Hubbard model, J. Phys. Soc. Jpn. {\bf 77}, 114713 (2008).
\bibitem{COCGSupp} S. Jin, A. Bulgac, K. Roche, G. Wlaz\l{}owski, 
Coordinate-Space Solver for Superfluid Many-Fermion Systems with Shifted Conjugate Orthogonal Conjugate Gradient Method, (2016), arXiv:1608.03711 [nucl-th].
\bibitem{GandolfiSM} S. Gandolfi, A. Gezerlis, and J.   Carlson, Neutron matter from low to high density, Annu. Rev. Nucl. Part. Sci. {\bf 65}, 303 (2015).
\bibitem{Avogadro2Supp} P. Avogadro, F. Barranco, R.A. Broglia, E. Vigezzi,
  Vortex-nucleus interaction in the inner crust of neutron stars, Nucl. Phys. A
  {\bf 811}, 378 (2008).
\bibitem{LambSM} H. Lamb, Hydrodynamics, Cambridge University Press (1975).
\bibitem{EpsteinBaymSM} R.I. Epstein and G. Baym, Vortex pinning in neutron stars, Ap. J. 328, 680 (1988).
\bibitem{MagierskiBulgacSM} P. Magierski, A. Bulgac, Nuclear hydrodynamics in the inner crust of neutron stars, Acta Phys. Polon. B35, 1203 (2004).	
\bibitem{MagierskiSM} P. Magierski, In-medium ion mass renormalization and lattice vibrations in the neutron star crust, Int. J. Mod. Phys. E13, 371 (2004).
\bibitem{CaoSM} L.G. Cao, U. Lombardo, amd P. Schuck, Screening effects in superffluid nuclear and neutron matter within Brueckner theory, Phys. Rev. C {\bf 74}, 064301 (2006).
\bibitem{UmarSM} A.S. Umar, V.E. Oberacker, C.J. Horowitz, P.-G. Reinhard, J.A. Maruhn, Swelling of nuclei embedded in neutron gas and consequnces for fusion, Phys. Rev. C {\bf 92}, 025808 (2015).
\bibitem{SevesoSM} S. Seveso, P. M. Pizzochero, F. Grill and B. Haskell, Mesoscopic pinning forces in neutron star crusts, MNRAS {\bf 455}, 3952 (2016).
\end{thebibliography}
\end{document}